\documentclass[12pt,aps,pra,preprint,groupedaddress,superscriptaddress]{revtex4}
\usepackage{graphicx}
\usepackage{amsmath,amssymb,amsfonts}
\usepackage{natbib}

\newcommand{\be}{\begin{equation}}
\newcommand{\ee}{\end{equation}}
\newcommand{\ba}{\begin{eqnarray}}
\newcommand{\ea}{\end{eqnarray}}

\begin{document}

\title{Linear response theory of activated surface diffusion with
interacting adsorbates}

\author{R. Mart\'{\i}nez-Casado}
\affiliation{Department of Chemistry, Imperial College London,
South Kensington, London SW7 2AZ, United Kingdom }

\author{A. S. Sanz}
\affiliation{Instituto de F\'{\i}sica Fundamental,
Consejo Superior de Investigaciones Cient\'{\i}ficas,
Serrano 123, 28006 Madrid, Spain}

\author{J. L. Vega}
\affiliation{Instituto de F\'{\i}sica Fundamental,
Consejo Superior de Investigaciones Cient\'{\i}ficas,
Serrano 123, 28006 Madrid, Spain}

\author{G. Rojas-Lorenzo}
\affiliation{Instituto Superior de Tecnolog\'{\i}as y Ciencias Aplicadas,
Ave. Salvador Allende, esq.\ Luaces, 10400 - La Habana, Cuba}
\affiliation{Instituto de F\'{\i}sica Fundamental,
Consejo Superior de Investigaciones Cient\'{\i}ficas,
Serrano 123, 28006 Madrid, Spain}

\author{S. Miret-Art\'es}
\affiliation{Instituto de F\'{\i}sica Fundamental,
Consejo Superior de Investigaciones Cient\'{\i}ficas,
Serrano 123, 28006 Madrid, Spain}

\date{\today}

\begin{abstract}
Activated surface diffusion with interacting adsorbates is analyzed
within the Linear Response Theory framework.
The so-called interacting single adsorbate model is justified by means
of a two-bath model, where one harmonic bath takes into account the
interaction with the surface phonons, while the other one describes the
surface coverage, this leading to defining a collisional friction.
Here, the corresponding theory is applied to simple systems, such as
diffusion on flat surfaces and the frustrated translational motion in
a harmonic potential.
Classical and quantum closed formulas are obtained.
Furthermore, a more realistic problem, such as atomic Na diffusion
on the corrugated Cu(001) surface, is presented and discussed within
the classical context as well as within the framework of Kramer's
theory.
Quantum corrections to the classical results are also analyzed and
discussed.
\end{abstract}

\maketitle


\section{Introduction}
\label{sec1}

The main purpose of spectroscopic experiments involving a probe and a
system at thermal equilibrium with a reservoir or thermal bath consists
of measuring the system response under the perturbation caused by the
probe \cite{gordon,mcquarrie}.
The intrinsic properties of matter can be extracted by analyzing this
response, which thus becomes a very important tool for physicists, who
can elucidate the microscopic structure and its dynamics through it.
Many times this response can be described by first order perturbation
theory and determined by the spectrum of the spontaneous fluctuations
of the reservoir, as established by the so-called
{\it fluctuation-dissipation} (FD) {\it theorem} \cite{kubo}.
Within this linear approximation, the probability per time unit that
the full system formed by probe and reservoir changes from the initial
state to the final state is given by the Fermi golden rule.
More specifically, the probe transition probability from an initial
state to a certain final state is given by the time Fourier transform
of the autocorrelation function associated with the operator defined
by the matrix element of the corresponding interaction Hamiltonian.
This type of studies is based on the work developed by van Hove
\cite{vanhove,vineyard}, who introduced the space-time correlation
function (namely the van Hove function), a generalization of the
well-known pair-distribution function from the theory of liquids,
as a tool to study the scattering of probe particles (slow neutrons)
by quantum systems consisting of interacting particles at thermal
equilibrium.
Within the Born approximation in scattering theory, the nature of
the scattered particles as well as the details of the system-probe
interaction potential are largely irrelevant, this essentially reducing
the scattering problem to a typical statistical mechanics problem
\cite{lovesey}.
The linear response function of a system consisting of interacting
particles, also known as {\it dynamic structure factor} or
{\it scattering law}, can be then related to the
spontaneous-fluctuation spectrum of such particles and expressed
in terms of particle density-density correlation functions
\cite{lovesey,hansen,bee}.
As is known, a complete description of this dynamics can also be
obtained through the linear response theory, where the FD theorem
is used to derive alternative response function expressions for the
dynamic structure factor.

From {\it quasielastic He atom scattering} (QHAS) experiments at low
energies very detailed information about defect and adsorbate dynamics
on surfaces can be obtained \cite{lahee,manson,hofmann,eli1}.
Manson and Celli \cite{manson} generalized van Hove's theory of
neutron scattering by crystals and liquids to atom surface scattering
within the transition matrix formalism.
Within this approach, small coverage of adsorbates and/or defects were
assumed in order to ignore both their interactions as well as multiple
scattering with He atoms.
Diffuse elastic and inelastic features were interpreted again through
the dynamical structure factor which is proportional to the observed
line shapes and is expressed in terms of the transfer of energy and
parallel momentum to the He atoms before and after the scattering
process.
The QHAS technique has also been applied to study surface diffusion on
metals with different types of atomic and molecular adsorbates, the
diffusion of Na adatoms (at different coverages) on Cu(001) being one
of the most extensively studied systems.
In the case of massive particles, where very large timescales are
involved, the dynamic structure factor can be expressed in terms of
the adsorbate positions, which has led to the so-called {\it single
adsorbate} model.
Within this approach, where low coverage is assumed, surface diffusion
(i.e., the adsorbate motion) is described by classical stochastic
trajectories of adsorbates issued from solving the standard Langevin
equation.
As in the case of Brownian-like particles, this equation encompasses
two contributions: (1) a (external) force arising from the
deterministic, phenomenological adiabatic potential describing the
adsorbate-substrate interaction at zero surface temperature; and (2) a
stochastic force (usually, a Gaussian white noise) accounting for the
vibrational effects induced by the temperature on the surface lattice
atoms (and therefore on the adsorbates).
The system dynamics is then obtained after solving the Langevin
equation \cite{graham,elis,eli1,vega1,vega2}, analyzing the results
derived from it in terms of the so-called {\it motional narrowing
effect} \cite{vega1,vega2} as well as Kramers' turnover theory
\cite{vega3,guantes1} and the dephasing theory \cite{guantes2}.
Within this scenario, it is usually assumed that He-adsorbate
interactions play no role on the surface dynamics, thus being
scarcely analyzed.

When the coverage is increased, the dynamic structure factor also
provides valuable information about the nature of the
adsorbate-adsorbate interaction, which should be included in the
corresponding theoretical studies.
In this way, pairwise potential functions accounting for the
adsorbate-adsorbate interactions are usually introduced into Langevin
molecular dynamics (LMD) simulations \cite{elis}.
Recently, an alternative procedure has been considered, where such
pairwise interactions are described by a purely stochastic model,
namely the {\it interacting single adsorbate} (ISA) model
\cite{ruth1,ruth2,ruth3,ruth4}.
This approach, also within the standard Langevin framework, is based
on both the theory of spectral-line collisional broadening developed by
van Vleck and Weisskopf \cite{vvleck-weiss} and the elementary kinetic
theory of gases \cite{mcquarrie}, and explains fairly well the
experimental broadening observed with increasing coverage \cite{elis}.
The standard Langevin equation is solved with two different,
non-correlated noise functions: (1) a Gaussian white noise accounting
for the surface friction, as before, and (2) a white shot noise
\cite{gardiner} replacing the pairwise interaction potential which
simulates the adsorbate-adsorbate collisions.
A double Markovian assumption is therefore considered (for the
interaction with the surface and for the interaction among adsorbates).
This assumption holds because, on the one hand, substrate excitation
timescales are much shorter than the timescales associated with the
adatom motions (the maximum frequency of the substrate excitation is
around 20-30~meV, while the characteristic vibrational frequency of
the adatom is around 4-6~meV).
On the other hand, the time between consecutive collisions (measured
through a collisional frequency, which would depend on temperature,
coverage or the adparticle mass) is typically much longer than the
effective time that a collisional effect may last (i.e., the time
two adsorbates may effectively be in physical contact).
Therefore, memory effects are not taken into account.
Just as the adsorbate-surface interaction is characterized by a
(surface) friction, the adsorbate-adsorbate interaction will also be
describable in terms of a {\it collisional friction}, which will vary
with coverage.
With this simple stochastic model, where the total friction is the
sum of the substrate friction and the collisional friction, a good
agreement with the experimental results for coverages up to 0.12,
approximately, has been obtained.
For higher coverage values, the model cannot be applied due to the
appearance of ordered structures, as it has been observed \cite{ellis}
experimentally between 0.12 and 0.16.
Recently, the collisional friction has been estimated from experiments
with benzene on graphite \cite{allison0}.
Although further investigation at a microscopic level and
first-principle calculations are needed, this simple stochastic model
at moderate coverages is able to provide a complementary view of
diffusion (through the quasielastic Q-peak) and low frequency
vibrational motions (through the frustrated translational T-mode
peak), at or around zero energy transfers (very long time dynamical
processes), respectively.
This could be understood because any trace of the true interaction
potential seems to be wiped out due to the relatively large number
of collisions taking place at very long times (the time scale for
the diffusion regime).
Actually, this purely stochastic model can be derived from a
microscopic classical Hamiltonian model characterized by two baths
associated with two independent collections of harmonic oscillators
\cite{german}, one related to the surface phonons and the other one
describing the presence of adsorbates, the source of the collisional
friction.
In this model, the coupling to low-lying electron-hole excitations
(electroninc friction) was not considered for simplicity.
One of the purposes in the present work is to extend this two-bath
model to quantum activated diffusion.

Surface diffusion processes are also studied experimentally by means
of the so-called {\it spin echo techniques}.
Thus, since the advent of He spin echo (HeSE) spectroscopy
\cite{allison1,allison2,allison3} and the improved signal in the
neutron spin echo (NSE) spectrometer \cite{farago,fouquet}, fast
diffusion processes are more accessible and allow us to to better
determine interaction potentials.
In both types of experiments, the observable is the so-called
{\it intermediate scattering function} or {\it polarization function},
which is the inverse time Fourier transform of the dynamic structure
factor.
This is a complex function whose real and imaginary parts can be
observed experimentally \cite{allison2,allison3}.
The intermediate scattering function is also
the space Fourier transform of the van Hove function, which, in general,
is also a complex-valued function.
The complex character of these functions can be understood as a
signature of the quantum nature of the diffusion dynamics.
In this regard, a quantum Markovian theory of surface diffusion for
interacting adsorbates has also been proposed recently \cite{ruth5}.
The imaginary part of the van Hove function is important at small
values of time or high temperatures (timescales of the order of $\hbar
\beta$, with $\beta = 1/k_B T$, $k_B$ being Boltzmann's constant).
This dynamical regime takes place  when the mean de Broglie wavelength
$\lambda_B = \hbar/\sqrt{2mk_BT}$ ($m$ is the adsorbate mass) is of
the order of or greater than typical interparticle distances.
For these timescales and distances, the adparticle positions are no
longer variables, but Heisenberg operators that do not commute at two
different times.
This theory thus tries to reconcile classical and quantum calculations
when no diffusion by tunneling is considered.

The purpose of this work is to provide an analysis of activated surface
diffusion with interacting adsorbates within the linear response theory
framework.
The theoretical framework of the aforementioned two-bath model is then
applied to simple systems, such as diffusion on flat surfaces and the
frustrated translational motion in a harmonic potential, obtaining
classical and quantum closed formulas.
Moreover, a more realistic problem, such as atomic Na diffusion on
on the corrugated Cu(001) surface, is presented and discussed within
the classical context as well as within the framework of Kramer's
theory, introducing quantum corrections to the classical results which
will be analyzed and discussed.
Note that an appropriate understanding of this surface dynamics is very
important, for diffusion is a preliminary step in more complicated
surface phenomena, such as heterogeneous catalysis, crystal growth,
lubrication, associative desorption, etc.
Furthermore, the QHAS and HeSE techniques can be considered as the
surface science analogue of the quasielastic neutron scattering
techniques, which has been widely and successfully applied to analyze
diffusion in bulk.

According to our purposes, we have organized this work as follows.
In Section~\ref{sec2}, we give a general overview of the Linear
Response Theory applied to activated surface diffusion.
In particular, a revision of the classical two-bath model is presented.
Regarding the quantum version of this model, a proposal and discussion
are also given in order to describe the stochastic trajectories of the
adsorbates at different coverages.
In Section~\ref{sec3}, applications to simple models (flat surfaces
and driven damped harmonic oscillator) as well as to more realistic
problems, as atomic Na diffusion on Cu(001), are analyzed within the
context of classical and quantum dynamics.
Kramers's turnover theory is also analyzed within the classical
context.
Finally, in Section~\ref{sec4} we summarize the conclusions derived
from this work as well as some future work.


\section{General theory for activated surface diffusion}
\label{sec2}


\subsection{Dynamic structure factor and intermediate scattering
function}
\label{sec2.1}

Space-time correlation functions \cite{yip91} can be used to describe
the decay of spontaneous thermal fluctuations at surfaces, being
central to the study of transport phenomena.
These functions are defined as the thermodynamic average of the product
of two dynamical variables, each one expressing the instantaneous
deviation from its corresponding equilibrium value at particular
points on the surface and time.
A complete description of the particle dynamics in a many-body system
is then reached when the behavior of the corresponding correlation
functions over the entire wavenumber range is studied.
This range splits into different characteristic regions, each one
associated with a different set of properties of the system.
In the case of scattering experiments, since the momentum and energy
transfers are the relevant quantities, any correlation-function theory
has to be developed necessarily in terms of such quantities.
Space-time correlation functions can also be used to describe the
response of a fluid under a weak, external perturbation.
Indeed, the reason why space-time correlation functions are central
quantities in transport phenomena in fluids is, precisely, because of
the equivalence between spontaneous fluctuation and linear response.

In general, a surface local dynamical variable is defined as
\begin{equation}
 \label{eq:dv}
 A ({\bf R},t)= \frac{1}{\sqrt{N}} \sum_{i=1}^N  a_i (t)
  \delta ({\bf R} - {\bf R}_i (t)) ,
\end{equation}
where $a_i (t)$ is any physical quantity and ${\bf R}_i (t) = (x_i (t),
y_i(t))$ is the time-dependent center of the position operator of the
adparticle on a two-dimensional surface.
The corresponding fluctuation is usually defined as $\delta A({\bf R},t)
= A({\bf R},t) - \langle A({\bf R},t) \rangle_{\beta}$, where the
average on a canonical ensemble is denoted by $\langle \ \cdot \
\rangle_{\beta}$.
The dynamical variable conserves if it satisfies a continuity equation
of the form
\begin{equation}
 \label{eq:ace}
  \dot{A} ({\bf R},t) =
   - \nabla_{\bf R} \cdot {\bf J}_A ({\bf R},t) ,
\end{equation}
where ${\bf J}_A$ is the {\it current} associated with the $A$ variable
and the dot over $A$ denotes the total time derivative.
Here, the dynamical variables of particular interest are the number
density,
\begin{equation}
 \label{eq:nd}
 \rho ({\bf R},t) = \frac{1}{\sqrt{N}} \sum_{i=1}^N
  \delta ({\bf R} - {\bf R}_i (t)) ,
\end{equation}
and the current density
\begin{equation}
 \label{eq:cd}
 {\bf J}({\bf R},t)= \frac{1}{\sqrt{N}} \sum_{i=1}^N
  {\bf v}_i (t) \delta ({\bf R} - {\bf R}_i (t)) ,
\end{equation}
where ${\bf v}_i$ accounts for the velocities of the $N$ adparticles.
The corresponding van Hove fluctuation density autocorrelation function
\cite{vanhove,lovesey} reads as
\begin{equation}
 \label{eq:vhcf}
  G(|{\bf R} - {\bf R}'|,t) = \Sigma \langle \delta \rho
  ({\bf R}',0) \delta \rho ({\bf R},t) \rangle_{\beta} ,
\end{equation}
and a similar expression holds for the current density.
The adparticle density is given by $\rho = N/\Sigma$, where $\Sigma$
is the surface area and the coverage is defined by $\theta = N/
N_{\rm max}$, with $N_{\rm max}$ being the maximum number of sites
in the $\Sigma$ area.
In most physical systems, correlation effects are negligible at large
space or time separation, the asymptotic limit being a simple product
of thermodynamically averaged quantities.

In analogy to scattering of slow neutrons by crystals and
liquids \cite{vanhove,vineyard,lovesey}, the observable magnitude
in QHAS experiments is the so-called {\it differential reflection
coefficient},
\be
 \frac{d^2 \mathcal{R} (\Delta {\bf K}, \omega)}{d\Omega d\omega}
  = n_d \mathcal{F} S(\Delta {\bf K}, \omega) .
 \label{eq:DRP}
\ee
This coefficient gives the probability that the He atoms (probe
particles) scattered from the interacting adsorbates on the surface
reach a certain solid angle $\Omega$ with an energy exchange $\hbar
\omega =E_f - E_i$ and wave vector transfer parallel to the surface
$\Delta {\bf K} = {\bf K}_f - {\bf K}_i$.
In Eq.~(\ref{eq:DRP}), $n_d$ is the concentration of adparticles;
${\mathcal F}$ is the {\it atomic form factor}, which depends on the
interaction potential between the probe atoms in the beam and the
adparticles on the surface; and $S(\Delta {\bf K},\omega)$ is the
dynamic structure factor, which gives, apart from other peaks, the
Q and T-mode peaks, also providing a complete information about the
dynamics and structure of the adsorbates through particle distribution
functions.
Experimental information about long distance correlations is obtained
from the dynamic structure factor when considering small values of
$\Delta {\bf K}$, while information on long time correlations is
provided at small energy transfers, $\hbar \omega$.

The pair distribution functions are given by means of the van
Hove or time-dependent pair correlation function $G({\bf R},t)$
\cite{vanhove}.
This function is related to the dynamic structure factor by a double
Fourier transform, in space and time, as
\be
 S(\Delta {\bf K}, \omega)
  = \frac{1}{2 \pi \hbar N} \int \! \! \! \int G({\bf R},t)
  e^{i(\Delta {\bf K} \cdot {\bf R} -\omega t)}
   \ \! d{\bf R} \ \! dt .
 \label{eq:DSF}
\ee
Given an adparticle at the origin at some arbitrary initial time,
$G({\bf R},t)$ represents the average probability to find a
particle (the same or another one) at the surface position
${\bf R} = (x,y)$ at a time $t$.
Thus, this function generalizes the well--known pair distribution
function $g({\bf R})$ from statistical mechanics \cite{mcquarrie,hansen}
by providing information about the interacting particle dynamics.

The position operators of the adsorbates are given, in general, by the
corresponding Heisenberg operators (defined for all $j=1, \cdots , N$
adparticles and time $t$),
\be
 {\bf R}_j (t) = e^{i H t /\hbar} {\bf R}_j e^{- i H t/\hbar} ,
 \label{eq:hei}
\ee
where $H$ is the Hamiltonian of the total system.
As mentioned above, the space Fourier transform of the $G$-function is
the intermediate scattering function,
\ba
 I(\Delta {\bf K}, t)
 & = & N \int \! \! \! \int G({\bf R},t)
  e^{i \Delta {\bf K} \cdot {\bf R}} \ \! d{\bf R} \nonumber \\
 & = & \frac{1}{N}\langle \rho_{\Delta {\bf K}}
 (t) \rho^\dag_{\Delta {\bf K}} (0) \rangle_{\beta} ,
 \label{eq:ISF}
\ea
where the $\rho_{\Delta {\bf K}}$ operator defined as
\be
 \rho_{\Delta {\bf K}} (t) = \sum_{j=1}^N
  e^{- i {\Delta {\bf K}} \cdot {\bf R}_j (t)}
   = \rho^\dag_{- \Delta {\bf K}} (t)
 \label{eq:rho1}
\ee
is the Fourier component of the adsorbate number-density operator,
\be
 \rho ({\bf R},t) = \sum_{j=1}^N \delta ({\bf R} - {\bf R}_j (t)) .
 \label{eq:rho2}
\ee
Thus, in (\ref{eq:ISF}) the brackets denote the ensemble average
over the trajectories associated with each adsorbate ${\bf R}_j(t)$.
The intermediate scattering function is the typical observable issued
from HeSE and NSE experimental techniques.
Taking into account the relations (\ref{eq:DSF}) to (\ref{eq:rho2}),
we note that the dynamic structure factor can be expressed in terms of
a density-density correlation function and determined by the spectrum
of the spontaneous fluctuations.
Moreover, the {\it static structure factor}, defined as
$S(\Delta {\bf K}, t=0)$, is related to $g({\bf R})$, which
describes the instantaneous correlation between adsorbates.

From relations (\ref{eq:nd}) and (\ref{eq:cd}), one finds by direct
differentiation in the reciprocal space the continuity or number
conservation equation,
\begin{equation}
\label{eq:ce} \frac{\partial \rho_{\Delta {\bf K}} (t)}{\partial t} = i
\Delta {\bf K} \cdot {\bf J}_{\Delta {\bf K}} (t) ,
\end{equation}
which inserted into (\ref{eq:ISF}), and taking into account
(\ref{eq:DSF}), we obtain the basic relation
\begin{equation}
\label{eq:sje}
 S(\Delta {\bf K}, \omega) = \frac{\Delta {\bf K}^2}{\omega^2}
  J_l (\Delta {\bf K}, \omega) ,
\end{equation}
where $J_l$ is the longitudinal (projected along the wave vector
transfer direction) current correlation function.
This relation is rigorous as far as the adparticles are not being
created or absorbed.

The dynamic structure factor $S(\Delta{\bf K}, \omega)$ is an even
function in frequency.
Therefore, all the odd moments vanish and the frequency moments
or frequency sum rules (static correlation functions) are
$\Delta {\bf K}$-dependent quantities,
\begin{equation}
 \label{eq:sr}
 \omega^{2n} (\Delta{\bf K}) = \frac{1}{2 \pi} \int
  \omega^{2n} S(\Delta {\bf K}, \omega) d\omega
  = (-1)^n \left [ \frac{\partial^{2n} I(\Delta {\bf K},t)}
     {\partial t^{2n}} \right ]_{t=0} ,
\end{equation}
which are obtained from a Taylor series expansion of the intermediate
scattering function at short times (or small distances).
In particular, the zeroth-order moment corresponds to the static
structure factor,
\begin{equation}
 \label{eq:sf}
  \omega^0 (\Delta {\bf K}) = S(\Delta {\bf K}) ,
\end{equation}
since it describes the average distribution of interparticle distances
on the surface; the second order moment is related to the thermal speed
$v_0 = \sqrt{2 k_B T/m}$, as
\begin{equation}
 \label{eq:v0}
 \omega^2 (\Delta {\bf K}) = \Delta {\bf K}^2 v_0^2 .
\end{equation}

Due to the quantum character of the different operators introduced
above, several comments are worth stressing.
First, $\rho_{\Delta {\bf K}} (t)$ and $\rho_{\Delta {\bf K}}^\dag (0)$
commute only at $t=0$.
Second, the system studied here is assumed to be stationary and,
therefore, the origin of time is arbitrary for the correlation function
associated with the density operators.
Third, the complex character of the corresponding correlation function
is a signature of the quantum dynamics of the interacting system.
Fourth, the $G$-function is also complex, but the dynamic structure
factor is real and positive definite because it represents a
cross-section.
More properties of the $\rho_{\Delta {\bf K}}(t)$ operator, the
$G$-function and the dynamic structure factor can be found in Lovesey's
book \cite{lovesey}.
And fifth, the so-called {\it detailed balance condition} can be
expressed as \cite{schofield}
\be
 S(\Delta {\bf K}, \omega) = e^{\hbar \omega \beta}
  S(- \Delta {\bf K}, - \omega) ,
 \label{eq:DBP}
\ee
which expresses that the probability that a He atom loses an energy
$\hbar \omega$ is equal to $e^{\hbar \omega \beta}$ times the
probability that a He atom gains an energy $\hbar \omega$.

After van Hove \cite{vanhove}, if $R_0$ is the range of the
$G$-function and $T_0$ its relaxation time, $\hbar / R_0$ and
$\hbar / T_0$ determine the orders of magnitude of average momentum
and energy transfers in the scattering process of the probe particles,
which for light masses display the observable recoil effect.
Thus, the time variation of $G$ affects the total scattering and
angular distributions only for a particle spending at least a time of
order $T_0$ over a correlation length $R_0$.
Moreover, if the mean de Broglie wavelength, $\lambda_B$, is
small compared to interadparticle distances or the range of
adsorbate-adsorbate interaction, no quantum effect will manifest in
the $G$-function, which deals with pairs of adparticles separated by
distances of the order of $R_0$.
Nevertheless, for small timescales ($t \ll T_0$ or $t \sim  \hbar
\beta)$), the dynamics entirely concentrates in a region of the order
of or less than $\lambda_B$, and quantum effects are noticeable.
At these distances, the adparticles can be considered as a
two-dimensional free gas.
The imaginary part of the $G$-function is greater at small values of
time.


\subsection{The Hamiltonian for the system and the thermal bath}
\label{sec2.2}


\subsubsection{The one-bath model}
\label{sec2.2.1}

In order to go a step further into the dynamics, we need to specify a
Hamiltonian as introduced in Eq.~(\ref{eq:hei}).
In surface diffusion, the full system+bath Hamiltonian is usually
written \cite{eli1} as
\begin{eqnarray}
 H & = & \frac{p_x^2}{2m} + \frac{p_y^2}{2m} + V(x,y) \nonumber \\
  & & + \sum_{i=1}^{N}
 \left[ \frac{p_{x_i}^2}{2 m_i}+ \frac{m_i}{2}
  \left( \omega_{x_i} x_i - \frac{c_{x_i}}
    {m_i \omega_{x_i}} \ \! x \right)^2  \right]
   \nonumber \\
  & & + \sum_{i=1}^{N}
 \left[ \frac{p_{y_i}^2}{2 m_i}+ \frac{m_i}{2}
  \left( \omega_{y_i} y_i - \frac{c_{y_i}}
  {m_i \omega_{y_i}} \ \! y \right)^2 \right] ,
\label{HCL1}
\end{eqnarray}
where $(p_x,p_y)$ and $(x,y)$ are the adparticle momenta and positions,
and $(p_{x_i},x_i)$ and $(p_{y_i},y_i)$ with $i=1, \cdots , N$ are the
momenta and positions of the bath oscillators (phonons), with mass
and frequency given by $m_i$ and $\omega_i$, respectively; phonons with
polarization along the $z$-direction are not considered.
The Hamiltonian was originally considered by Magalinskii \cite{maga}
and Caldeira and Leggett \cite{caldeira}, who used it for weak and
strong dissipation (a general discussion about the Hamiltonian
(\ref{HCL1}) can be found in Weiss' book \cite{weiss}).
In surface diffusion, $V(x,y)$ is in general a periodic function
describing the surface corrugation at zero temperature.

The harmonic frequencies of the bath modes and the coupling
coefficients are expressed in terms of spectral densities, defined
\cite{caldeira} as
\begin{equation}
 J_i(\omega) = \frac{\pi}{2}
 \sum_{j=1}^N \frac{c_{i_j}^2}{m_j \omega_{i_j}^2}
  \left[ \delta (\omega - \omega_{i_j}) \right] .
 \label{SD1}
\end{equation}
with $i=x,y$.
This enables the passage to a continuum model.
The associated friction functions are defined through the cosine
Fourier transform of the spectral densities as
\be
\eta_i (t) = \frac{2}{\pi m} \int_0^{\infty} d \omega
\frac{J_i (\omega)}{\omega} \cos \omega t ,
\label{fric}
\ee
with $i=x,y$.
For Ohmic friction, $\eta_i (t) = 2 \eta_i \delta (t)$ where $\eta_i$
is a constant and $\delta (t)$ is the Dirac delta function.
In this model, the noise is shown to be white when Ohmic friction
is assumed.
The paradigm of white noise is the Gaussian white noise.
Dealing with large systems (the surface seen as a thermal bath) where
the number of collisions between substrate and adsorbate is very
high, the fundamental theorem of probability theory, namely the central
limit theorem, assures that the fluctuations of the bath will be
Gaussian distributed.
Diffusion can then be described by a Brownian-type motion involving a
continuous Gaussian stochastic process.
In virtue of the FD theorem, such fluctuations can be related to the
friction coming mainly from surface phonons: the phonon friction.
Electronic friction due to low-lying electron-hole pair excitations are
usually neglected in most of cases.


\subsubsection{The two-bath model}
\label{sec2.2.2}

The diffuse elastic intensity of the He atoms scattered at large angles
away from the specular direction provides very detailed information on
the mobility of adsorbates on surfaces.
Based on the transition matrix formalism, Manson and Celli \cite{manson}
proposed a quantum diffuse inelastic theory for small and intermediate
coverages of adsorbates on the surface by ignoring multiple scattering
effects of He atoms. The dynamical structure factor is then
obtained by assuming all the crystal vibrational modes ($N$)
and point-like scattering centers ($M$) satisfying the harmonic
approximation with a given frequency distribution function. Therefore,
following the same type of arguments, we could assume two independent,
uncorrelated baths to describe diffusion of interacting adsorbates.
As before, the first bath consists of the surface phonons.
Meanwhile, the second bath is formed by $M$ adsorbates which,
obviously, changes with the surface coverage given by experimental
conditions \cite{german}.

For a two-bath model (for a given coverage), we take one adsorbate as
the tagged particle or system, while the remaining ones constitute the
second bath descibed by $M$ harmonic oscillators.
In this way, the corresponding total Hamiltonian reads as \cite{german}
\begin{eqnarray}
 H & = & \frac{p_x^2}{2m} + \frac{p_y^2}{2m} + V(x,y) \nonumber \\
  & & + \sum_{i=1}^{N}
 \left[ \frac{p_{x_i}^2}{2 m_i}+ \frac{m_i}{2}
  \left( \omega_{x_i} x_i - \frac{c_{x_i}}
    {m_i \omega_{x_i}} \ \! x \right)^2  \right]
   \nonumber \\
  & & + \sum_{i=1}^{N}
 \left[ \frac{p_{y_i}^2}{2 m_i}+ \frac{m_i}{2}
  \left( \omega_{y_i} y_i - \frac{c_{y_i}}
  {m_i \omega_{y_i}} \ \! y \right)^2 \right]
  \nonumber \\
     & & + \sum_{j=1}^{M}
 \left[ \frac{{\bar p}_{x_j}^2}{2 {\bar m}_j} + \frac{{\bar m}_j}{2}
  \left( {\bar \omega}_{x_j} {\bar x}_j
  - \frac{d_{x_j}}{{\bar m}_j {\bar \omega}_{x_j}} \ \! x \right)^2
   \right]
  \nonumber \\
  & & + \sum_{j=1}^{M}
 \left[ \frac{{\bar p}_{y_j}^2}{2 {\bar m}_j} + \frac{{\bar m}_j}{2}
  \left( {\bar \omega}_{y_j} {\bar y}_j
   - \frac{d_{y_j}}{{\bar m}_j {\bar \omega}_{y_j}} \ \! y \right)^2
   \right] ,
\label{HCL2}
\end{eqnarray}
where the barred magnitudes label the same quantities as in the
one-bath model, but now referring to a bath of $M$ adsorbates, which
are also considered as a collection of harmonic oscillators.
The $c_{k_j}$ and $d_{k_j}$ coefficients, with $k = x, y$, give the
coupling strengths between the adsorbate and the substrate phonons or
other adsorbates, respectively.
The spectral density for the two baths is defined analogously to the
one-bath model,
\begin{equation}
  J_i(\omega) = \frac{\pi}{2}
 \sum_{j=1}^N \frac{c_{i_j}^2}{m_j \omega_{i_j}^2}
  \left[ \delta (\omega - \omega_{i_j}) \right]
   + \frac{\pi}{2}
 \sum_{j=1}^M \frac{d_{i_j}^2}{{\bar m}_j {\bar \omega}_{i_j}^2}
  \left[ \delta (\omega - {\bar \omega}_{i_j}) \right] ,
 \label{SD2}
\end{equation}
but now it is split into two sums, one spectral density due to the
surface phonons and the other one due to adsorbates.
In a similar way, the friction functions are defined as in
Eq.~(\ref{fric}) but now the spectral density is given by (\ref{SD2}).
The friction function is also split into two terms,
one due to the phonons, $\gamma (t)$, and the other one due to the
presence of adsorbates, $\lambda (t)$. This last friction function
could be interpreted like a collisional friction. Again, for Ohmic
friction, $\eta_i (t) = 2 (\gamma + \lambda) \delta (t)$,
where $\gamma$ and $\lambda$ are constant and $\delta (t)$
is the Dirac delta function.

Extension of the one-bath model to two baths is carried out to describe
collisions among adsorbates when the coverage is increased up to a
certain value.
Again, if the friction is assumed to be Ohmic, the noise will be white.
Adsorbate collisions can be seen as discrete events.
It is well known that an appropriate way to model this type of noise is
by considering a Poisson process, being generally designated as a white
shot noise.
As has been shown elsewhere \cite{ruth2,ruth3}, this white shot noise
can be obtained as a limiting case of a color noise.
It is also clear that when the diffusion regime is reached,
the discrete Poisson process becomes a continuous Gaussian process. The
introduction of the second bath allows us to describe adsorbate
collisions by a new friction coefficient, the collisional friction. If
the two baths are not correlated, the corresponding noises are also not
correlated and this is the key point of the ISA model. The total
friction coefficient, the sum of the phonon and collisional frictions,
is related to the fluctuations of both baths through the FD theorem.
This sum of frictions has recently been estimated by the He spin
echo technique \cite{allison0,allison3}.

Before concluding this subsection, we would like to emphasize that if
an external  driving force is also present (He atoms are sampling the
surface to explore the motion of adparticles) a new extra term
should be added to the total Hamiltonian, (\ref{HCL1}) or
(\ref{HCL2}) (see Section~{\ref{sec3}).


\subsection{The Langevin equation in the two-bath model}
\label{sec2.3}

The next task is to eliminate the environmental (two-baths)
degrees of freedom, which leads to a damped equation of motion
of the system coordinates. If the Heisenberg picture of quantum
mechanics is used, where the time evolution of a given operator
$A$ is given by
\be
{\dot A} = \frac{i}{\hbar} [H,A] ,
\label{heis}
\ee
a generalized Langevin equation (GLE) for each coordinate
of the system is obtained, reading as follows
\begin{equation}
 m \ddot{x}(t) + m \int_0^t \eta_x (t-t') \ \!
 \dot{x}(t') \ \! dt' + \frac{\partial V(x,y)}{\partial x} =
  N_x (t)
 \label{GLEx}
\end{equation}
and
\begin{equation}
 m \ddot{y}(t) + m \int_0^t \eta_y (t-t') \ \!
 \dot{y}(t') \ \! dt' + \frac{\partial V(x,y)}{\partial y} =
  N_y (t) ,
 \label{GLEy}
\end{equation}
where the associated friction functions are defined through the cosine
Fourier transform of the spectral densities given by Eq.~(\ref{SD2}),
\be
\eta_i (t) = \frac{2}{\pi m} \int_0^{\infty} d \omega
\frac{J_i (\omega)}{\omega} \cos \omega t ,
\label{fric2}
\ee
with $i=x,y$.
Due to the splitting of the spectral density (\ref{SD2}), note that the
friction function also splits into two terms, one due to the phonons,
$\gamma (t)$, and the other one due to the presence of adsorbates,
$\lambda (t)$.

The inhomogenity in Eqs.~(\ref{GLEx}) and (\ref{GLEy}) represents a
fluctuating force which depends on the initial position of the system
and initial positions and momenta of the oscillators of each bath; a
generalization of the one bath model \cite{weiss}.
The fluctuating force in each direction can again be split into two
sums, one due to the phonons and the other due to adsorbates. As
both baths are assumed to be uncorrelated, the same property holds
for the two noises. For each noise and each cartesian component,
it can be easily shown that its equilibrium (canonical ensemble)
expectation value with respect to the heat bath
including the corresponding bilinear coupling to the system vanishes.
On the contrary, the noise autocorrelation function (each cartesian
component and each bath) is a complex quantity because in general
it does not commute at different times.
In the classical limit $\hbar \rightarrow 0$, each
noise correlation reduces to $m k_B T \eta_i (t)$, with $i=x,y$.
For Ohmic friction, i.e., delta correlated, we have white noises.
In the quantum case \cite{weiss,ingold}, and also for Ohmic friction,
the imaginary part of each noise function is a step function and
its real part goes with ${\rm csch}^2 (\pi t / \hbar \beta)$.
Thus, at zero surface temperature, the noise is still correlated even
for long time (it decays algebraically like $t^{-2}$) in contrast to
the classical case. These facts give rise to important differences with
respect to the classical case such as, for example, the noise and
the system coordinates are correlated instead of being zero.
In order to simplify the theory, we will consider only classical noise
but keeping in mind that our quantum results (see Section~\ref{sec3})
will be only valid for not too low surface temperatures. In the
QHAS and HeSE experimental techniques used for fast diffusion,
the lowest attainable surface temperature is around
50--100~K. In a future work, quantum noise and tunneling will be
considered in surface diffusion problems at very low temperatures.

Thus, if $\eta (t) = 2 \eta \delta (t) = 2(\gamma + \lambda)\delta (t)$
(Ohmic friction), Eqs.~(\ref{GLEx}) and (\ref{GLEy}) reduce to two
coupled standard Langevin equations (Markovian approximation) (the
delta function counts only one a half when the integration is carried
out from zero to infinity) \cite{ruth1,ruth2,ruth3},
\begin{equation}
 m \ddot{\bf R} = - m \eta \dot{\bf R} - {\bf F}({\bf R}) ,
  + \delta {\bf N} ,
 \label{LME}
\end{equation}
which is the basis of the ISA model, and where $\delta {\bf N}$ is
given by the sum of two noncorrelated noises: the lattice (thermal)
vibrational effects and the adsorbate-adsorbate collisions, which are
simulated by a Gaussian white noise ($G$) and a shot white noise ($S$)
(Poissonian distributed), respectively.
Thus, for each degree of freedom, we have $\delta N (t) = \delta N_G
(t) + \delta N_S (t)$.
Let us remark that a Poissonian distribution behaves as a Gaussian
distribution for very long times.
The Langevin equation is then solved for one single particle
in presence of two noises.

For a good simulation of a diffusion process, one
has to consider very long times in comparison to the timescales
associated with the friction caused by the surface or to the typical
vibrational frequencies observed when the adsorbates keep moving
inside a surface well. This means that there will be a considerably
large number of collisions during the time elapsed in the propagation,
and therefore that, at some point, the past history of the adsorbate
could be irrelevant regarding the properties we are interested in.
This memory loss is a signature of a Markovian dynamical regime, where
adsorbates have reached what we call the {\it statistical limit}.
Otherwise, for timescales relatively short, the interaction is not
Markovian and it is very important to take into account the effects
of the interactions on the particle and its dynamics (memory effects).
The diffusion of a single adsorbate is thus modeled by a series of
random pulses within a Markovian regime (i.e., pulses of relatively
short duration in comparison with the system relaxation) simulating
collisions among adsorbates.
In particular, we describe these adsorbate-adsorbate collisions by
means of a white shot noise as a limiting case of a colored shot
noise \cite{gardiner}, as mentioned above.
This interaction is, therefore, described in terms of the collisional
friction, which depends on the surface coverage.
Thus, the ISA model essentially consists of solving
the standard Langevin equation with two noise sources and frictions:
a Gaussian white noise accounting for the friction with the substrate
and a white shot noise characterized by a collisional friction
simulating the adsorbate-adsorbate collisions. This way of simulating
the interaction among adsorbates reduces the dynamical problem to
the diffusion of a single adsorbate (like in the SA approximation)
and the $N$ factor appearing in the $S$ or $I$ functions,
Eqs.~(\ref{eq:DSF}) and (\ref{eq:ISF}), has not to be considered.

Finally, the surface coverage $\theta$ and $\lambda$ or collisional
friction can be related in a simple manner.
In the elementary kinetic theory of transport in gases \cite{mcquarrie}
diffusion is proportional to the mean free path $\bar{l}$, which is
proportionally inverse to both the density of gas particles and the
effective area of collision when a hard-sphere model is assumed.
For two-dimensional collisions the effective area is replaced by an
effective length (twice the radius $\rho$ of the adparticle) and the
gas density by the surface density $\sigma$.
Accordingly, the mean free path is given by
\begin{equation}
 \bar{l} = \frac{1}{2 \sqrt{2} \rho \sigma} .
 \label{mfp}
\end{equation}
Taking into account the Chapman-Enskog theory for hard spheres, the
self-diffusion coefficient can be written as
\begin{equation}
 D = \frac{1}{6 \rho \sigma} \ \! \sqrt{\frac{k_B T}{m}} .
 \label{d-mfp}
\end{equation}
Now, from Einstein's relation, and taking
into account that $\theta = a^2 \sigma$ for a square surface lattice
of unit cell length $a$, we finally obtain
\begin{equation}
 \lambda = \frac{6\rho}{a^2} \ \! \sqrt{\frac{k_B T}{m}} \ \! \theta .
 \label{theta}
\end{equation}
Therefore, given a certain surface coverage and temperature, $\lambda$
can be readily estimated from (\ref{theta}).
Notice that when the coverage is increased by one order of magnitude,
the same holds for $\lambda$ at a given temperature. Notice that
$\lambda$ could also be considered as a phenomenological parameter,
just like the substrate friction $\gamma$.


\subsection{Linear response functions}
\label{sec2.4}

The dynamic structure factor can also be related to the linear response
function of the system \cite{lovesey},
\be
\phi (\Delta {\bf K}, t) = \frac{i}{\hbar N} \langle
[\rho_{\Delta {\bf K}} (t), \rho_{\Delta {\bf K}}^\dag] \rangle ,
\label{eq:LRF}
\ee
through the FD theorem, expressed as
\be
S(\Delta {\bf K}, \omega) = \frac{1}{2 \pi i} [ 1 + n(\omega) ]
 \int^\infty_{- \infty} e^{i \omega t} \ \phi (\Delta {\bf K}, t) \ dt ,
\label{eq:FDT}
\ee
where $1 + n(\omega) = [1 - \exp(- \hbar \omega \beta)]^{-1}$
is the detailed balance factor, with
$n(\omega)$ the Boltzman factor.
Equation~(\ref{eq:FDT}) relates the spectrum of spontaneous
fluctuations, $S(\Delta {\bf K}, \omega)$, to the dissipation
part of the response function.
Time derivatives of the response function are related to the Heisenberg
equation of motion (\ref{heis}), of the $\rho_{\Delta {\bf K}}$
operator and moments of the dynamic structure factor involve nested
commutators to evaluate them.
The $\phi$-function is a causal function because it can not
be defined before the external perturbation has been switched on.
For the scattering with He atoms, the perturbation starts at
$- \infty$ and ends at $+ \infty$ and typically has a bell shape
(see Section~{\ref{sec3}).

In (\ref{eq:FDT}), the time Fourier transform of $\phi$ defines
a {\it generalized susceptibility function}
$\chi (\Delta {\bf K},\omega)$ and, therefore, can be reexpressed as
\be
S(\Delta {\bf K}, \omega) = - i [ 1 + n(\omega)]
\chi (\Delta {\bf K}, \omega) .
\label{eq:FDT-S}
\ee
This susceptibility is complex and the real and imaginary parts are
related through the well-known Kramers-Kronig or dispersion relations
\cite{lovesey}.
On the other hand, the time derivative of the linear response function
is related to the so-called {\it relaxation function} as follows
\be
 R_{\phi} (\Delta {\bf K}, t) = \int_t^{\infty} dt'
 \phi (\Delta {\bf K}, t') ,
 \label{eq:RF}
\ee
which describes the relaxation of the density after the external
perturbation (He atoms) has been switched off. Physically, $R_{\phi}$
goes to zero as $t \rightarrow \infty$ and at $t=0$ this function
coincides with the isothermal susceptibility. Thus, the dynamic
structure factor can again be reexpressed in terms of the relaxation
function as
\be
S(\Delta {\bf K}, \omega) = [ 1 + n(\omega) ] \, \omega \,
{\tilde R}_{\phi} (\Delta {\bf K}, \omega) ,
\label{eq:FDT-R}
\ee
where ${\tilde R}_{\phi}(\Delta {\bf K}, \omega)$ is the time Fourier
transform of the relaxation function.

Finally, the dynamic structure factor can also be expressed in terms
of the Green function which is directly related to the linear response
function, generalized susceptibility and the relaxation function
\cite{lovesey},
\be
 S(\Delta {\bf K}, \omega) = - \frac{1}{\pi \hbar}
  \left[ 1 + n(\omega) \right]
  {\rm Im} \ \! G (\Delta {\bf K}, \omega) ,
\label{Green1}
\ee
where
\be
 {\rm Im} G (\Delta {\bf K}, \omega) =
   - \hbar \ \! {\rm Im}\ \! \chi (\Delta {\bf K}, \omega) ,
\label{green2}
\ee
with ${\rm Im}\ \! G (\Delta {\bf K}, \omega)$ being the imaginary part
of the Green function.


\section{Applications}
\label{sec3}

In general, an exact, direct calculation of $I(\Delta {\bf K},t)$
or $S(\Delta {\bf K},\omega)$ is difficult to carry out due to the
noncommutativity of the adparticle position operators at different
times, which obey a Langevin-Markovian equation, here described by
(\ref{LME}), where the friction is assumed Ohmic and the interaction
potential is not separable.
However, for certain simple cases, close formulas can be easily
obtained.

The product of the two exponential operators in (\ref{eq:ISF}) can
be evaluated according to a special case of the Baker-Hausdorff
theorem (the {\it disentangling theorem}). If $A$ and $B$ are two
operators then  $e^A e^B =  e^{[A,B]/2} e^{A+B}$, which only
holds when the corresponding commutator is a c-number.
Thus, (\ref{eq:ISF}) reads now as \cite{ruth5}
\begin{equation}
 I (\Delta {\bf K},t) =
  I_1 (\Delta {\bf K},t) I_2 (\Delta {\bf K},t) ,
 \label{eq3}
\end{equation}
which is a product of two quantum intermediate scattering functions
$I_j (\Delta{\bf K},t)$ with $j=1,2$ associated with the exponentials
of the commutator $[A,B]$ and $A+B$, respectively.
If we identify \cite{ruth5} the operators $A$ and $B$ as $A= i \Delta
{\bf K} \cdot {\bf R}(0)$ and $B= - i \Delta {\bf K} \cdot {\bf R}(t)$,
the factor $I_1$ involving their commutator will depend on the
character of the dynamics; for classical dynamics, this factor
is one.
The second factor can also be written as follows
\be
 I_2(\Delta {\bf K},t) =
 \langle e^{- i \Delta {\bf K} \cdot (\hat{\bf R}(0)- \hat{\bf R}(t))}
 \rangle
 = \langle e^{- i \Delta K \int_0^t \hat{v}_{\Delta {\bf K}} (t') dt']}
  \rangle \simeq e^{- \Delta K^2 \int_0^t (t-t') C_v(t') dt'}
 \label{eq7}
\ee
within the so-called Gaussian approximation and where
$C_v(t) = \langle {v}_{\Delta {\bf K}} (t) {v}_{\Delta {\bf K}}
(0) \rangle$  is the velocity autocorrelation function along the
direction given by $\Delta {\bf K}$ or the longitudinal direction.
Equation~(\ref{eq7}) is exact if the velocity operator gives rise to
a Gaussian stochastic process.
In the cases we are going to analyze below, we will discuss the
factorization given by (\ref{eq3}) in more detail.


\subsection{Diffusion on flat surfaces}
\label{sec3.1}

In the case of diffusion on flat or very low corrugated surfaces,
where the role of the adiabatic adsorbate-substrate interaction
potential is negligible and only the action of the thermal phonons
and surrounding adsorbates are relevant, one can assume $V (x,y)
\approx 0$. Thus, the stochastic
single-particle trajectories ${\bf R}(t)$ running on the surface obey
the following Langevin-Markovian equation (\ref{LME})
\begin{equation}
 m \ddot{\bf R}(t) = - m \eta \dot{\bf R}(t) + \delta {\bf N}(t) ,
 \label{eq4}
\end{equation}
where $\delta {\bf N}(t) = \delta {\bf N}_G (t) + \delta {\bf N}_S (t)$
is the two-dimensional fluctuation of the total noise acting on the
adparticle.

He atoms are usually the probe particles and it is generally assumed
that they do not influence the surface dynamics, that is, their
influence can be considered a perturbation. Some considerations are
in order about the driving force or external perturbation.
First, the attractive part of the
interaction potential does not play an important role in vibrational
excitation problems, and it can usually be neglected. And second, the
incident energy for the incoming particles is large compared to the
vibrational excitation of the adsorbate (the frustrated translational
or rotational modes which are of very low frequency).
A semiclassical description of the
projectile-adsorbate interaction allows for an estimate of the
collisional time and the duration of energy transfer. Thus, if $r$ is
the distance between the He atom and the center of mass of the adsorbate
and their interaction is accepted to be exponentially repulsive
\cite{david}, $V_e (r) = A \exp[- \alpha ' r]$, it can be easily
shown that the external force can be expressed as $F_e (t)
= C \ \! {\rm sech}^2 \alpha ' t$ with $C= \alpha ' v_i \sin \theta_i /2$ where
$v_i$ and $\theta_i$ are the incident velocity and angle, respectively.
The parameter $\alpha '$ gives the rate of energy exchange between the
translational (He atoms) and frustrated translational
(adsorbates) motions. The hyperbolic
function has the physically correct behavior at the asymptotic limits
($t \rightarrow \pm \infty$), and it is maximum at $t=0$ where the
closest distance to the adsorbate is reached (bell shape).
A similar expression can be obtained for the external force if
instead of a pure repulsive interaction a Morse interaction is used.

Due to the fact that for a flat surface no direction is priveleged,
and if the adsorbate motion is driven by the external force $F_e(t)$,
we have from Eq.~(\ref{eq4}) that
\be
\langle \ddot {x}(t) \rangle_{\beta} + \eta \langle \dot{x}(t)
\rangle_{\beta} =  \frac{1}{m} F_e(t) .
\label{eq71}
\ee
Within the linear response framework we could write the
particular solution of the differential equation as
\be
\langle {\tilde x}(t) \rangle_{\beta} = \int_{- \infty}^t ds
\phi (t-s) F_e(s) ,
\label{eq72}
\ee
or, after Fourier transforming,
\be
\langle \tilde{R} (\omega) \rangle_{\beta} = \chi (\omega)
\tilde{F}_e (\omega) ,
\label{eq73}
\ee
where
\be
 \tilde{F}_e (\omega) = \frac{A \omega}{\alpha'^2} \ \!
  {\rm csc} \left( \frac{\pi \omega} {2 \alpha'} \right) .
\label{eq74}
\ee
Now, if we assume that $\omega / \alpha'$ keeps close to unity during
the interaction along time, the cosecant function will be also close to
unity and the dynamic susceptibility will be that of a free adparticle
on a flat surface,
\be
\chi (\omega) = \frac{1}{m} \frac{1}{- \omega^2 - i \eta \omega} ,
 \label{eq8}
\ee
which is exact whenever an Ohmic friction $\eta$ is assumed and
any direction given by $\Delta K$ is considered.
This expression of the dynamic susceptibility is valid for both
the classical and quantum case.


\subsubsection{Classical dynamics}
\label{sec3.1.1}

The adparticle motion can then be regarded as quasi-free since it is
not ruled by a potential, but only influenced by the two stochastic
forces.
Within this regime, the velocity is a Gaussian stochastic process and
the velocity autocorrelation function in any direction (since there is
no priveleged direction) is given by Doob's theorem \cite{ruth3},
\begin{equation}
 \mathcal{C}_v(t) = \frac{k_B T}{m} \ \! e^{- \eta t} .
 \label{corrGM2}
\end{equation}
The expression for the intermediate scattering function resulting
from (\ref{eq7}), where $I(\Delta K,t) = I_2(\Delta K,t)$ is
\begin{equation}
 I(\Delta K,t) = \exp \left[- \chi^2
   \left( e^{- \eta t} + \eta t - 1 \right) \right] ,
 \label{eq:IntSGM}
\end{equation}
where the so-called {\it shape parameter} $\chi$
\cite{jcp-elliot,ruth3} is defined as
\begin{equation}
 \chi^2 \equiv \langle v_0^2 \rangle \Delta K^2 / \eta^2 .
 \label{eq:chi2}
\end{equation}
From this relation we can obtain both the mean free path $\bar{l}
\equiv \tau \sqrt{ \langle v_0^2 \rangle }$ and the
self-diffusion coefficient $D\equiv \tau\langle v_0^2\rangle$
({\it Einstein's relation}).
When the coverage increases, the collisions among adsorbates are also
expected to increase, and so $\lambda$ and therefore $\eta$.
As can be easily shown, (\ref{eq:IntSGM}) displays a Gaussian decay
at short times that does not depend on the particular value of $\eta$
(ballistic motion), while at longer times it decays exponentially with
a rate given by $\eta^{-1}$.
Thus, with $\eta$, the decay of the intermediate scattering function
becomes slower.

The above described effects can be quantitatively understood by means
of the expression of the dynamic structure factor obtained analytically
from (\ref{eq:IntSGM}),
\ba
 S (\Delta K, \omega) & = & \frac{e^{\chi^2}}{\pi \eta} \ \!
  \chi^{-2\chi^2} \ \! {\rm Re} \left\{ \chi^{-2i\omega/\eta}
   [ \tilde {\Gamma} (\chi^2 + i\omega/\eta)
    - \tilde {\Gamma} (\chi^2 + i\omega/\eta, \chi^2)] \right\}
 \nonumber \\
 & = & \frac{e^{\chi^2}}{\pi}
  \sum_{n=0}^\infty \frac{(-1)^n \chi^{2n}}{n!}
   \frac{(\chi^2 + n) \eta}{\omega^2 + [ (\chi^2 + n) \eta]^2} \ \! .
 \label{dsf1}
\ea
Here, the $\tilde{\Gamma}$ symbol in the first line denotes both the
Gamma and incomplete Gamma functions (depending on the corresponding
argument), respectively.
As can be noted in the high friction limit, (\ref{dsf1}) becomes a
Lorentzian function, its {\it full width at half maximum} (FWHM) being
$\Gamma = 2\eta\chi^2$, which approaches zero as $\eta$ increases
(narrowing effect).
This is in sharp contrast to what one could expect --- as the frequency
between successive collisions increases one would expect that the line
shape gets broader (effect of the pressure in the spectral lines of
gases).
The physical reason for this effect could be explained as follows.
As $\eta$ increases particle's mean free path decrease and
therefore correlations are lost more slowly.
In the limit case where friction is such that the particle remains in
the same place, the van Hove function becomes a $\delta$-function, the
intermediate scattering function remains equal to one and the dynamic
structure factor consists of a $\delta$-function at $\omega = 0$.
Conversely, in the low friction limit the line shape is given by a
Gaussian function, whose width is $\Gamma = 2\sqrt{2\ln 2}
\sqrt{k_B T/m} \ \! \Delta K$, which does not depend on $\eta$.
This is the case for a two--dimensional {\it free gas}
\cite{toeprl,ruth6}.
This gradual change of the line shapes as a function of the friction
and/or the parallel momentum transfer leading to a change of the shape
parameter $\chi$ is known as the
{\it motional narrowing effect} \cite{vega1,vega2}.
Notice that, in our approach, friction is related to the coverage. Thus,
at higher coverages a narrowing effect is predicted for a flat surface
\cite{ruth1}.


\subsubsection{Quantum dynamics}
\label{sec3.1.2}

In the Heisenberg representation, Eq.~(\ref{eq4}) still holds, its
formal solution being
\begin{equation}
 {\bf R} (t) =  {\bf R} (0) + \frac{{\bf P} (0)}
 {m \eta} \ \! \Phi (\eta t) +  \frac{1}{m \eta} \int_0^t
 \Phi (\eta t - \eta t') \delta {\bf N} (t') dt' ,
 \label{eq5}
\end{equation}
where ${\bf P} (0)$ is the initial adparticle momentum operator and
$\Phi(x)=1-e^{-x}$. Then, the commutator between ${\bf R} (0)$ and
${\bf R} (t)$ is obtained from Eq.~(\ref{eq5}) obtaining a c-number.
Then, the factor $I_1$,
by assuming a classical noise as previously mentioned, reads as
\cite{ruth5}
\begin{equation}
 I_1 (\Delta {\bf K},t) = \exp \left[ \frac{i \hbar \Delta {\bf K}^2}
 {2 \eta m} \ \! \Phi(\eta t) \right] =
  \exp \left[ \frac{i E_r}{\hbar} \frac{\Phi(\eta t)}{\eta} \right] ,
 \label{eq6}
\end{equation}
where $E_r = \hbar^2 \Delta {\bf K}^2/2m$ is the {\it adsorbate recoil
energy}. As is apparent, the argument of the exponential function
becomes less important as the adparticle mass and the total friction
increase. The time-dependence only comes from  $\Phi(\eta t)$.
At short times ($\lesssim \hbar \beta$),
$\Phi(\eta t) \approx \eta t$ and the argument of $I_1$ becomes
independent of the total friction, thus increasing linearly with time.
On the other hand, in the asymptotic time limit, this argument
approaches a constant phase.

In order to calculate the $I_2$ factor we start from Eq.~(\ref{eq4})
describing an adparticle with mass $m$ moving on a flat surface in
presence of more adsorbates.
The dynamic susceptibility is also given by Eq.~(\ref{eq8}) and its
time behavior by
\be
\chi (t) = \frac{2}{m \eta} e^{- \eta t/2} \sinh (\eta t/2)
\Theta (t)
\label{eq9}
\ee
where $\Theta (t)$ is the step function due to causality.
The FD theorem allows us to express the equilibrium position
autocorrelation function, $C_x (t) =  \langle x(t) x(0) \rangle$, in
terms of the imaginary part of the dynamic susceptibility and, after
Fourier transforming, we find
\be
C_x (t) = \frac{\hbar}{\pi m} \int_{- \infty}^{+ \infty} d \omega
\frac{\eta \omega}{\omega^4 + \eta^2 \omega^2}
\frac{e^{- i \omega t}}{1 - e^{- \beta \hbar \omega}} .
 \label{eq10}
\ee
From the relations
\be
\frac{1}{1 - e^{- \beta \hbar \omega}} =
\frac{1}{2} + \frac{1}{2} \ \! {\rm coth} \left( \beta \hbar \omega /2
  \right)
 \label{eq11}
\ee
and
\be
 {\rm coth} (\beta \hbar \omega /2) = \frac{2}{\beta \hbar \omega}
  \left( 1 + 2 \sum_{n=1}^{\infty} \frac{\omega^2}{\nu_n^2 + \omega^2}
    \right) ,
 \label{eq12}
\ee
where the so-called Matsubara frequencies are defined by
\be
\nu_n = \frac{2 \pi n}{\hbar \beta} ,
 \label{eq13}
\ee
we can decompose the correlation function as $C_x(t) = S_x(t) + i
A_x(t)$, i.e., into its symmetric and antisymmetric parts,
respectively.
For $t > 0$, these functions read as
\be
S_x (t) = - \frac{1}{m \beta \eta} \left( t\ \! {\rm sign} [ t ]
  + \frac{1}{\eta} e^{- \eta t} \right) + \frac{2}{\beta m}
 \sum_{n=1}^{\infty} \frac{\eta e^{- \nu_n t} - \nu_n e^{- \eta t}}
  {\nu_n  (\eta^2 - \nu_n^2)}
 \label{eq14}
\ee
and
\be
A_x(t) = - \frac{\hbar}{2 \eta m} \left( 1 - e^{- \eta t} \right) ,
 \label{eq15}
\ee
which can be trivially related to (\ref{eq9}) by the FD theorem.
In (\ref{eq14}), the sign function of the real number $t$
is defined as follows: $+1$ for $t > 0$ and $-1$ for $t < 0$.
Now, since
\be
C_v (t) = - \frac{d^2}{dt^2} C_x(t) ,
 \label{eq16}
\ee
then
\be
C_v (t) =  \left( \frac{1}{\beta m} - i \frac{\hbar \eta}{2 m}
\right) e^{- \eta t} - \frac{2 \eta}{\beta m} \sum_{n=1}^{\infty}
\frac{\nu_n e^{- \nu_n t} - \eta e^{- \eta t}}{\eta^2 - \nu_n^2} ,
 \label{eq17}
\ee
where the real part is identical to \ref{corrGM2}) except for the
infinite sum of the {\it Matsubara frequencies}.
Quantum effects are important at low surface temperatures, the long
time behavior being mainly determined by the first term of the
Matsubara series.
In such cases, relaxation is no longer governed only by the damping
constant \cite{weiss,ingold}.

Finally, (\ref{eq17}) is substituted into (\ref{eq7}) in order to
obtain the factor $I_2$,
\be
I_2(\Delta {\bf K},t) =  e^{- \Delta K^2 \int_0^t (t-t')
C_v(t') dt'} = e^{- \Delta K^2 [f(t) + g(t)]} ,
 \label{eq18}
\ee
where the time-dependent functions $f(t)$ and $g(t)$ are given by
\be
f(t) = \left( \frac{1}{m \beta \eta^2} - i \frac{\hbar}{2 m \eta}
 \right) [ e^{- \eta t} + \eta t -1]
 \label{eq19}
\ee
and
\be
g(t) = \frac{2}{m \beta} \sum_{n=1}^{\infty}
\frac{\nu_n e^{- \eta t} - \eta e^{- \nu_n t} + \eta - \nu_n}
{\nu_n (\eta^2 - \nu_n^2)} .
 \label{eq20}
\ee
The total intermediate scattering function (\ref{eq3}) can be then
expressed as
\be
I(\Delta {\bf K},t) = e^{- \chi^2 [\alpha^* \eta t -
\Phi (\eta t)]} e^{- \Delta K^2 g(t)} ,
 \label{eq21}
\ee
with $\chi = \Delta K^2 \langle v_0^2 \rangle / \eta^2$ and
$\alpha = 1 + i \hbar \beta / 2$, the thermal square velocity being
$\langle v_0^2 \rangle = 1/m \beta$. The recoil energy
is included in the imaginary part of the product
$\chi^2 \alpha^*$, which disappears when $\hbar \rightarrow 0$.
Equation~(\ref{eq18}) is slightly different to that obtained elsewhere
\cite{ruth5}, where the factor $I_2$ was treated classically,
\begin{equation}
 I_2 (\Delta {\bf K},t) = e^{- \chi^2 [ \eta t - \Phi(\eta t) ] } ,
 \label{eq22}
\end{equation}
and, therefore, the total intermediate scattering function (\ref{eq3})
can be expressed as
\begin{equation}
 I (\Delta {\bf K},t) = e^{\alpha \chi^2} \,
  e^{- \chi^2 [ \eta t + \alpha \Phi(\eta t) ] } .
 \label{eq23}
\end{equation}
Equation~(\ref{eq21}) is the generalization of the intermediate
scattering function for the quantum motion of interacting adsorbates
in a flat surface.
The dependence of this function on $\Delta {\bf K}^2$ through the shape
parameter is the same as in the classical theory \cite{ruth3}.
No previous information about the velocity autocorrelation
function is needed.
However, classically, the intermediate scattering function
is usually obtained from Doob's theorem, which states that the velocity
autocorrelation function for a Gaussian, Markovian stationary
process decays exponentially with time \cite{risken}.
The ballistic or free-diffusion regime and the diffusive regime
are apparent from (\ref{eq23}). The first one is dominant at very
low times, $\eta t \ll 1$, and  the second one at very long times,
$\eta t \gg 1$.

The  diffusion coefficient can be obtained from
\be
D = \lim_{t \rightarrow \infty} \int_0^{t} C_v (t') dt' .
 \label{eq24}
\ee
Thus, from Eqs.~(\ref{eq17}) and (\ref{eq19}), the diffusion
coefficient is the real part of the complex number given by
\be
D = \frac{k_B T}{m \eta} - i \frac{\hbar}{2 m} ,
\label{eq25}
\ee
which coincides with Einstein's law for the classical case
(insuring that the adparticle velocity distribution becomes
Maxwellian asymptotically). The same result is reached from the mean
square displacement, $\langle ({\bf R}(t) - {\bf R}(0))^2 \rangle$,
which takes into account only the symmetric part of the position
autocorrelation function. Quantum fluctuations (the Matsubara
frequencies) do not affect this result at low temperatures except
the time limit to which the mean square displacement (MSD) is linear
with time may become very large. At zero temperature, $D$ is also
zero and the MSD is no longer linear with time. The infinite sum of
Matsubara frequencies determines now the long time limit behavior.
As previously mentioned, the limit to zero surface temperatures is
questionable if in the commutator we neglect the correlation between
the noise and the coordinate system.


\subsection{Diffusion on harmonic potentials}
\label{3.2}

The harmonic model is an appropriate working model to understand
the bound motion inside the wells of a
corrugated surface and, therefore, to also understand the behavior
associated to the T-mode, which comes precisely from the oscillating
behavior undergone by the adparticle when the diffusional motion is
temporarily frustrated.
Now if we again assume that $\omega / \alpha'$ keeps close to unit
during the interaction along time, the cosecant function will be
close to one and the dynamic susceptibility will be that of an
adparticle subject to a one-dimensional harmonic potential
\be
\chi (\omega) = \frac{1}{m} \frac{1}{- \omega^2 - i \eta \omega
+ \omega_0^2} ,
 \label{eq8b}
\ee
$\omega_0$ being the frequency of the harmonic oscillator. This
expression is exact whenever an Ohmic friction $\eta$ is assumed.
This expression of the dynamic susceptibility is valid for both
the classical and quantum cases.


\subsubsection{Classical dynamics}
\label{sec3.2.1}

In contrast with the case of a dynamics where $V(x,y)$ does not play a
relevant role, we can devise a particle fully trapped within a
harmonic potential well.
Thus, for a harmonic oscillator, the behavior of the adparticle becomes
very apparent when looking at the corresponding velocity autocorrelation
function, which reads \cite{risken,vega1,ruth3} as
\begin{equation}
 C_v(t) =
  \frac{k_B T}{m} \ \! \frac{\omega_0}{\bar{\omega}} \ \!
   e^{- \eta t/2} \cos (\bar{\omega} t + \delta_1) ,
 \label{corrHO2}
\end{equation}
with
\begin{equation}
 \bar{\omega} \equiv \sqrt{ \omega_0^2 - \frac{\eta^2}{4} } ,
\end{equation}
and $\tan \delta_1 = \eta / 2 \bar{\omega}$.
Note that (\ref{corrGM2}) can be easily recovered after some algebra
in the limit $\omega_0 \to 0$ from (\ref{corrHO2}).

The only information about the structure of the lattice is found
in the shape parameter through $\Delta K$ [see Eq.~(\ref{eq:chi2})].
When large parallel momentum transfers are considered, both the
periodicity and the structure of the surface have to be taken into
account.
Consequently, the shape parameter should be changed for different
lattices.
The simplest model including the periodicity of the surface is
that developed by Chudley and Elliott \cite{elliott}, who proposed
a master equation for the pair-distribution function in space and
time assuming instantaneous discrete jumps on a two-dimensional
Bravais lattice. Very recently, a generalized shape parameter based on
that model has been proposed to be \cite{jcp-elliot}
\be
 \chi_l (\Delta {\bf K}) \equiv \sqrt{\frac{\Gamma_{\nu}
 (\Delta {\bf K})}{2 \eta}} ,
 \label{eq11b}
\ee
where, within our approach, $\gamma$ is substituted by $\eta$.
Here, $\Gamma_{\nu} (\Delta {\bf K})$ represents the inverse of the
correlation time and is expressed as
\be
 \Gamma_{\nu} (\Delta {\bf K}) = \nu
 \sum_{\bf j} P_{\bf j} \ \! [1 - \cos({\bf j} \cdot \Delta {\bf K})] ,
 \label{eq9b}
\ee
$\nu$ being the total jump rate out of an adsorption site and
$P_{\bf j}$ the relative probability that a jump with a displacement
vector ${\bf j}$ occurs.

Substituting now (\ref{corrHO2}) into (\ref{eq7}) leads to the
following expression for the intermediate scattering function
\begin{equation}
 I(\Delta K,t) =
  \exp \left\{ - \frac{\chi_l^2 \eta^2}{\bar{\omega} \omega_0}
   \left[ \cos \delta_1 - e^{-\eta t/2}
    \cos (\bar{\omega} t - \delta_1) \right] \right\} .
 \label{isfho}
\end{equation}
The argument of this function displays an oscillatory behavior around
a certain value with the amplitude of the oscillations being
exponentially damped.
This translates into an also decreasing behavior of the intermediate
scattering function, which also displays oscillations around the
asymptotic value.
This means that after relaxation the intermediate scattering function
has not fully decayed to zero unlike the case of absence of a
potential.
Again, in the limit $\omega_0 \to 0$, (\ref{isfho}) approaches
(\ref{eq:IntSGM}).

In order to obtain an analytical  expression for the dynamic structure
factor, it is convenient to express (\ref{isfho}) as \cite{ruth3}
\ba
 I(\Delta K, t) & = & e^{-\chi_l^2 f(\bar{\omega}, t)}
 \nonumber \\
 & = & e^{-\chi_l^2 A_1}
  \sum_{m,n=0}^\infty \frac{(-1)^{m+n}}{m!\ \! n!}
  \ \! \chi_l^{2(m+n)} A_3^m A_4^n
  e^{i(m-n)\delta_1} e^{-(m+n) \eta t/2} e^{i(m-n) \bar{\omega} t} ,
 \label{isfho2}
\ea
where
\be
 f(\bar{\omega}, t) \equiv
  A_1 + A_3 e^{i\delta_1} e^{-(\eta/2 - i\bar{\omega})t}
  + A_4 e^{-i\delta_1} e^{-(\eta/2 + i\bar{\omega})t} ,
\ee
with
\ba
 A_1 & = & \frac{\omega_0}{\bar{\omega}}
 \frac{\eta^2 \{2 (\eta/2) \bar{\omega} \sin \delta_1
  + [\bar{\omega}^2 - (\eta/2)^2] \cos \delta_1\}}
   {[(\eta/2)^2 + \bar{\omega}^2]^2} ,
 \\
 A_3 & = & \frac{\omega_0}{\bar{\omega}}
  \frac{\eta^2}{2(\eta/2 - i\bar{\omega})^2} ,
 \\
 A_4 & = & \frac{\omega_0}{\bar{\omega}}
  \frac{\eta^2}{2(\eta/2 + i\bar{\omega})^2} ,
\ea
where the coefficients $A_i$ has been put in terms of $\eta$,
$\bar{\omega}$ and $\delta_1$.
From (\ref{isfho2}), it is now straightforward to derive an expression
for the dynamic scattering factor, which results
\ba
 S(\Delta K, \omega) & = & \frac{e^{-\chi_l^2 A_1}}{\pi}
  \sum_{m,n=0}^\infty \frac{(-1)^{m+n}}{m!\ \! n!}
  \ \! \chi_l^{2(m+n)} A_3^m A_4^n e^{i(m-n)\delta_1}
 \nonumber \\ & & \qquad \quad \times
   \frac{(m+n) \eta/2}{[\omega - (m-n)\bar{\omega}]^2
    + [(m+n) \eta/2]^2} .
 \label{dsfho2}
\ea
For a harmonic oscillator, there is no diffusion and, therefore,
(\ref{dsfho2}) is only valid when $m \neq n$.
All the Lorentzian functions contributing to (\ref{dsfho2}) are due to
the creation and annihilation events of the T mode.
These Lorentzians are characterized by a width given by
$\Gamma = (m+n) \eta/2$, which increases with $\eta$.
This broadening (proportional to $\eta$) undergone by the dynamic
structure factor is thus contrary to the narrowing effect observed
in the case of a flat surface.
It can be assigned to the confined or bound motion displayed by the
particle ensemble when trapped inside the potential wells.
Hence, in order to detect broadening of the line shapes in surface
diffusion experiments, adparticles must spend some time confined inside
potential wells, since the broadening will be induced by the presence
of temporary vibrational motions.


\subsubsection{Quantum dynamics}
\label{3.2.2}

The formal solution of Eq.~(\ref{LME}) is given by
\begin{equation}
 {\bf R} (t) =  {\bf R} (0) + \frac{{\bf P} (0)}
 {m \eta} \ \! \Phi (\eta t) +  \frac{1}{m \eta} \int_0^t
 \Phi (\eta t - \eta t') [{\bf F} ({\bf R}(t')) + \delta
{\bf N} (t')] dt',
 \label{eq5b}
\end{equation}
where the force ${\bf F}$ is given by Hooke's law and
${\bf P} (0)$ is the initial adparticle momentum operator and
$\Phi(x)=1-e^{-x}$. The presence of the adiabatic force
introduces an additional commutator, $[{\bf R}_0,{\bf F}({\bf R}(t))] =
(i\hbar)\partial {\bf F}({\bf R}(t))/\partial {\bf P}_0$,
where the dependence of the adiabatic force on the initial state
$({\bf R} (0), {\bf P} (0))$ is through ${\bf R}(t)$ which is again
negligible in a quantum Markovian framework \cite{ruth5}.
Thus the factor $I_1$ is the same as for a flat surface,
Eq.~(\ref{eq6}).

In order to calculate the $I_2$ factor we again start from
Eq.~(\ref{LME}).
The dynamic susceptibility is also given by (\ref{eq8b}) and its
time behavior by
\be
\chi (t) = \frac{1}{m {\bar {\omega}}} e^{- \eta t/2}
\sin {\bar {\omega}} t \Theta (t) ,
\label{eq9bb}
\ee
where $\Theta (t)$ is again the step function due to causality.
According to the FD theorem, as before, the equilibrium position
autocorrelation function can be expressed as
\be
C_x (t) = \frac{\hbar}{\pi m} \int_{- \infty}^{+ \infty} d \omega
\frac{\eta \omega}{(\omega^2 - \omega_0^2)^2 + \eta^2 \omega^2}
\frac{e^{- i \omega t}}{1 - e^{- \beta \hbar \omega}}
 \label{eq10b}
\ee
whose symmetric and antisymmetric parts are given, for $t > 0$, as
\be
S_x (t) = \frac{e^{- \eta t /2}}{m \beta {\bar {\omega} \omega_0^2}}
  [ \bar {\omega} \cos {\bar {\omega}} t + (\gamma/2)
 \sin \bar {\omega} t ] - \frac{2}{\beta m} \sum_{n=1}^{\infty}
  \frac{e^{- \nu_n t}} {(\eta / 2 - \nu_n)^2 + {\bar {\omega}^2}}
 \label{eq14b}
\ee
and
\be
 A_x(t) = - \frac{\hbar}{2 m {\bar {\omega}}} e^{- \eta t /2}
 \sin {\bar {\omega}} t ,
 \label{eq15b}
\ee
respectively, and the velocity autocorrelation function by
\be
C_v (t) = \frac{\omega_0}{m \beta {\bar {\omega}}}e^{- \eta t /2}
 \cos ({\bar {\omega}} t + \delta_1)
 + \frac{2}{\beta m} \sum_{n=1}^{\infty}
 \frac{\nu_n^2 e^{- \nu_n t}}{(\eta / 2 - \nu_n)^2 -
 {\bar {\omega}^2}}
 - \frac{i\hbar \omega_0^2}{2 m \bar {\omega}} e^{- \eta t /2}
 \cos (\bar {\omega}t + \delta_2) ,
 \label{eq17b}
\ee
with $\tan \delta_2 = \eta {\bar {\omega}} / \omega_0^2$.
Again, the real part is the same as in the classical case except
the Matsubara series. The same considerations about the surface
temperature in the quantum regime can be done as before.
The $I_2$ factor can be expressed as in (\ref{eq18}) through the
functions
\begin{eqnarray}
f(t) & = & \frac{t}{m \beta \bar {\omega}} e^{- \eta t /2}
 \sin \bar {\omega}t + \frac{2}{\beta m} \sum_{n=1}^{\infty}
 \frac{\nu_n^2 e^{- \nu_n t}}{(\eta / 2 - \nu_n)^2 -
 {\bar {\omega}^2}} \nonumber \\
 & & + \frac{i\hbar t}{2 m}   e^{- \eta t /2}
  \left[ (1 - \gamma^2 / \omega_0^2) (e^{\eta t /2}
 - \cos \bar{\omega}t)
 + \frac{\gamma^3 - 3 \gamma \omega_0^2}
 {2 \omega_0 \bar{\omega}} \sin \bar{\omega} t \right]
 \label{eq171}
\end{eqnarray}
and
\begin{eqnarray}
g(t) & = & \frac{1}{m \beta \omega_0^2 \bar {\omega}}
 \left\{ \bar {\omega} - e^{- \eta t /2} [ \bar {\omega}
 \cos \bar {\omega} t + (\omega_0^2 t + \gamma /2)
 \sin \bar {\omega} t ] \right\} \nonumber \\
 & & - \frac{2}{\beta m} \sum_{n=1}^{\infty}
 \frac{1 - e^{- \nu_n t} (\nu_n t + 1)}{(\eta / 2 - \nu_n)^2 -
 {\bar {\omega}^2}}
 + \frac{i\hbar \omega_0^2}{2 m \bar {\omega}}
 \left[ g_0 + g_1(t) + g_2(t) \right] ,
 \label{eq172}
\end{eqnarray}
with
\ba
g_0 & = & \frac{\bar {\omega} \eta}{4 \omega_0^7} (\eta^3 + 2 \omega_0
 \eta^2 - 2 \omega_0^2 \gamma - 4 \omega_0^3) , \\
g_1(t) & = & \frac{e^{- \eta t /2}}{\omega_0^6}
  [{\bar {\omega}} \omega_0^2 (\omega_0^2 - \eta^2 ) t
 + 2 \omega_0^2 \eta {\bar {\omega}} - \eta^3 {\bar {\omega}}]
 \cos \bar {\omega} t , \\
g_2(t) & = & \frac{e^{- \eta t /2}}{\omega_0^6}
 [(\eta / 2) \omega_0^2 (3 \omega_0^2 - \eta^2) t +
 4 \omega_0^2 \eta^2 - \omega_0^4 - \eta^4/2 ] \sin \bar{\omega} t .
\ea
The total intermediate scattering function will be the product of
the factors $I_1$ and $I_2$ given by (\ref{eq6}) and (\ref{eq18}),
taking into account (\ref{eq171}) and (\ref{eq172}), respectively.


\subsection{Diffusion on corrugated surfaces}
\label{sec3.3}


\subsubsection{Classical dynamics}
\label{sec3.3.1}

The broadening of the diffusion line shapes has been shown to be
produced by the temporary trapping \cite{ruth1}. We start this
subsection by considering  a general velocity autocorrelation function
which keeps the same functional form of Eq.~(\ref{corrHO2}),
but whose parameters do not hold the same
relations as those characterizing a harmonic oscillator \cite{vega1},
that is,
\begin{equation}
 C_v(t) = \frac{k_B T}{m} \ \! e^{- \tilde{\eta} t}
  \cos (\tilde{\omega}t + \tilde{\delta}) ,
 \label{corrg}
\end{equation}
where the values of the parameters $\tilde{\eta}$, $\tilde{\omega}$ and
$\tilde{\delta}$ are obtained from a fitting to the numerical results
issued from solving the standard Langevin equation with periodic
boundary conditions (\ref{LME}).
From Eq.~(\ref{corrg}) one easily reaches the corresponding
expression for the intermediate scattering function, \cite{ruth3}
\ba
 I(\Delta K, t) & = & e^{-\chi_l^2 \tilde{f}(\tilde{\omega},t)}
 \nonumber \\
 & = & e^{-\chi_l^2 \tilde{A}_1 - \chi_l^2 \tilde{A}_2 t}
  \sum_{m,n=0}^\infty \frac{(-1)^{m+n}}{m! \ \! n!} \ \!
  \chi_l^{2(m+n)} \tilde{A}_3^m \tilde{A}_4^n
  e^{i(m-n)\tilde{\delta}} e^{-(m+n)\tilde{\eta}t}
   e^{i(m-n)\tilde{\omega}t} ,
 \label{isfg}
\ea
which is analogous to (\ref{isfho2}), and where
\be
 \tilde{f}(\tilde{\omega},t) \equiv \tilde{A}_1 + \tilde{A}_2 t
  + \tilde{A}_3 e^{i\tilde{\delta}} e^{-(\tilde{\eta}
   - i\tilde{\omega})t}
  + \tilde{A}_4 e^{-i\tilde{\delta}} e^{-(\tilde{\eta}
   + i\tilde{\omega})t} ,
\ee
and
\ba
 \tilde{A}_1 & = & \frac{\tilde{\eta}^2
  [2 \tilde{\eta} \tilde{\omega} \sin \tilde{\delta}
  + (\tilde{\omega}^2 - \tilde{\eta}^2) \cos \tilde{\delta})}
  {(\tilde{\eta}^2 + \tilde{\omega}^2)^2} ,
 \\
 \tilde{A}_2 & = &
  \frac{\tilde{\eta}^2 (\tilde{\eta} \cos \tilde{\delta}
   - \tilde{\omega} \sin \tilde{\delta})}
  {\tilde{\eta}^2 + \tilde{\omega}^2} ,
 \\
 \tilde{A}_3 & = & \frac{\tilde{\eta}^2}
  {2(\tilde{\eta} - i\tilde{\omega})^2} ,
 \\
 \tilde{A}_4 & = & \frac{\tilde{\eta}^2}
  {2(\tilde{\eta} + i\tilde{\omega})^2} .
\ea
Unlike the case of the harmonic oscillator, notice now that there is
a new linear dependence on time in $\tilde{f}$ because of the
parameters $\tilde{\eta}$, $\tilde{\omega}$ and $\tilde{\delta}$
are time-independent.
This leads to an exponentially decaying factor in (\ref{isfg}), which
accounts for the diffusion and that makes the intermediate scattering
function to vanish at asymptotic times.
In this sense, the intermediate scattering function can be considered
as containing both phenomena, diffusion and low vibrational motions.
This effect is better appreciated in the dynamic structure factor,
\ba
 S(\Delta K, \omega) & = & \frac{e^{-\chi_l^2 \tilde{A}_1}}{\pi}
  \sum_{m,n=0}^\infty \frac{(-1)^{m+n}}{m!\ \! n!}
  \ \! \chi_l^{2(m+n)} \tilde{A}_3^m \tilde{A}_4^n e^{i(m-n)\delta}
 \nonumber \\
 & & \qquad \times
   \frac{\chi_l^2 \tilde{A}_2 + (m+n) \tilde{\eta}}
   {[\omega - (m-n)\tilde{\omega}]^2
    + [\chi^2 \tilde{A}_2 + (m+n) \tilde{\eta}]^2} .
 \label{dsfg}
\ea
This general expression clearly shows that both motions (diffusion and
oscillatory) cannot be separated at all.
The Q--peak is formed by contributions where $m=n$, for which each
partial FWHM is given by
\be
 \Gamma_Q = \chi_l^2 \tilde{A}_2 + 2 \, m \, \tilde{\eta}.
\ee
Analogously, the T-mode peaks come from the sums with $n \neq m$ and
each partial FWHM is given by
\be
 \Gamma_T = \chi_l^2 \tilde{A}_2 + (m + n) \, \tilde{\eta}.
\ee
If the Gaussian approximation is good enough,
the value of $\tilde{\eta}$ will not be too different from
the nominal value of $\eta$ and, therefore, both peaks
will display broadening as $\eta$ increases.
This is a very remarkable result since a relatively simple model,
as the one described here, can explain the corresponding experimental
broadenings observed with coverage. Thus, broadening arises from the
temporary confinement of the adparticles inside potential wells along
their motion on the surface \cite{ruth1}. The problem of the
experimental deconvolution has been already discussed elsewhere
\cite{jcp-elliot}. Here we would like only to mention that using this simple
model, such deconvolutions would be more appropriate in order to
extract useful information about diffusion constants and jump mechanisms.
Finally, as mentioned before, the motional narrowing effect will
govern the gradual change of the whole line shape as a function of
the friction or, equivalently, the coverage, the parallel momentum
transfer and the jump mechanism.


\subsubsection{Classical dynamics. The Kramer's turnover framework}
\label{sec3.3.2}

The theory of activated surface diffusion in one dimension was
developed \cite{melnikov,eli3} from Kramers' solution to the problem
of escape from a metastable well. \cite{kramers,peter} The underlying
dynamics is assumed well described by the Langevin equation provided
that the reduced barrier height is of the order of $\sim 3$ or
higher, the energy loss to the bath of trajectories close to the
barrier top is given by classical mechanics and the potential at the
barrier top is approximately parabolic. It has been shown that
Kramers' based theory with finite barrier correction terms can
then be replaced by Langevin numerical simulations \cite{guantes1}.

The starting point is a kinetic equation for the stationary flux of particles
exiting each well at either barrier. This flux is affected by the rate
of particles exiting the $j$th well and those arriving at the well
from the two neighboring wells, $j+1$ and $j-1$. Here we are going to give the
main analytical expressions derived from Kramers' theory, more details
can be found elsewhere. \cite{eli3,guantes1} A central quantity
in the theory is the reduced average energy loss $\delta$ to the bath as the
adatom traverses from one barrier to the next. For a single cosine potential
as that considered here with barrier height $V^{\ddagger} = 2 V_0$,
the energy loss is given by
\be
\delta = \frac{8 V_0 \gamma'}{k_B T \omega_0} ,
\label{delta}
\ee
where $\omega_0 = 2 \pi \sqrt{V_0 / m a^2}$ is the harmonic frequency
of oscillation near the well bottom, $m$ is the mass of the adatoms
and $a$ is the unit cell length. Since typically many experiments or
calculations are carried out under conditions of large reduced barrier
heights, $\delta$ can be unity or even larger, even though the damping
constant is rather small.

In the moderate to strong friction limit where the rate limiting step
is spatial diffusion (sd) across the barrier, the rate of the escape
from the well in both directions is given by the Kramers-Grote-Hynes
formula \cite{kramers,grote}
\be
\Gamma_{sd}=\frac{\Lambda^{\ddagger}}{\omega^{\ddagger}}
\frac{\omega_{0}}{\pi} \exp[-V^{\ddagger}/ k_B T] ,
\label{sd}
\ee
where the Kramers-Grote-Hynes prefactor is
\be
\frac{\Lambda^{\ddagger}}{\omega^{\ddagger}}=
\sqrt{1+\frac{\eta^{2}}{4\omega^{\ddagger}}}-
\frac{\eta}{2\omega^{\ddagger}} ,
\label{pref}
\ee
which has been generalized to two baths and
appears as a renormalization taking into account recrossings,
since we are working implicitly in normal mode coordinates for the
diffusing particle and the two baths. Finally, for the partial
rates one finds
\be
\Gamma_{j} = -\frac{\Gamma_{sd}}{\pi}\int_{0}^{2\pi}d \Delta K
\sin^{2} \left(\Delta K/2\right) \cos(j \Delta K)
 \exp \left\{ \frac{2}{\pi}\int_{0}^{\pi/2}dx
\ln \left[ \frac{1-P^{2}(x)}{1+P^{2}(x)\cos(\Delta K)} \right] \right\} ,
\ee
where the function $P(x)$ is defined as
\be
 P(x) = \exp \left[-\frac{\delta}{4\cos^{2}(x)} \right] .
\ee
and $\Delta K$ is written dimensionless. The rate of escape from the
zeroth well,
\be
 \kappa = - \Gamma_0,
\ee
and the relative probability or a jump of length $j$ is given by the
probability of being trapped at the $j$th well,
\be
 P_j = \frac{\Gamma_j}{\kappa} .
\ee
For a one-dimensional periodic potential, the diffusion coefficient is
related to the escape rate by
\be
 D = \frac{1}{2} \kappa \langle a^2 \rangle = \frac{1}{2} a^2
  \sum_j^{\infty} j^2 \Gamma_j ,
\ee
where $\langle a^2 \rangle$ is the mean square path length.
The diffusion coefficient can then be expressed in close form as
\be
 D = D_{sd} \Upsilon^{-1} \exp \left\{\frac{2}{\pi}\int_{0}^{\pi/2}dx
  \ln[1+P(x)] \right\} ,
\ee
with $D_{sd} = (1 / 2 a^2) \Gamma_{sd}$ the diffusion coefficient in
the spatial diffusion regime, and $\Upsilon$ the depopulation factor
for the metastable well first given by Melnikov \cite{melnikov},
\be
 \Upsilon =
  \exp\left\{\frac{2}{\pi}\int_{0}^{\pi/2}dx\ln[1-P(x)]\right\} .
\ee

In analogy to the Chudley-Elliott model \cite{elliott,ruth2}, an
analytical expression for the FWHM of the dynamic structure factor can
also be obtained by imposing a master equation for the intermediate
scattering function. One easily sees that if the dynamic structure
factor has the ubiquitous Lorentzian shape, the FWHM is given by
\be
 \Gamma(\Delta K) = 4 \Gamma_{sd} \sin^{2}
  \left(\frac{\Delta K}{2}\right)
 \exp \left\{\frac{2}{\pi}\int_{0}^{\pi/2}dx\ln \left[
 \frac{1-P^{2}(x)} {1+P^{2}(x)-2P(x) \cos(\Delta K)} \right] \right\} .
\ee
This equation is important in the sense that assuming the validity of
Kramers' model and the master equation approach, it allows for a
direct comparison with the experimental data and/or Langevin
numerical simulations and therefore an estimation of the spatial
diffusion rate $\Gamma_{sd}$ and the energy loss $\delta$. From these
parameters and their temperature dependence, one can further infer the
barrier height, the friction coefficient and the barrier frequency.

In order to solve Eq.~(\ref{LME}) we have used the velocity-Verlet
algorithm, which is commonly applied when dealing with stochastic
differential equations \cite{allen}.
For the average calculations shown here a number of 10,000
trajectories is sufficient for convergence along the $[100]$ direction.
The initial conditions are chosen such that the velocities are
distributed according to a Maxwell--Boltzmann velocity distribution
at a temperature $T$ (i.e., the ensemble is initially thermalized),
and the positions such that they cover the extension of a single
unit cell of the potential model used (see below).
Regarding the dynamical parameters, we have used  the same parameters
previously chosen for the Na atom on a Cu(001) surface, that is,
$\gamma = 0.1 \omega_0$, where $\omega_0$ is obtained from the
harmonic approximation assumed near the well bottom with a barrier
of $V_0 = 41.4$ meV. The mass and radius of the
adparticles are those of a Na atom, since this
adsorbate has been widely used in QHAS experiments over the Cu(001)
surface (non-separable potential) \cite{ruth4}.
As for the coverage, $\theta = 1$ corresponds
to one Na atom per Cu(001) surface atom or, equivalently,
$\sigma = 1.53 \times 10^{19}$~atom/cm$^2$;
$a = 2.557$~\AA\ is the unit cell length; and $\rho = 2$~\AA\
has been used for the atomic radius. Once the surface temperature
and the coverage are fixed, the corresponding
$\lambda$ value is obtained from Eq.~(\ref{theta}).

In Fig.~\ref{fig1} jump rates calculated in ps$^{-1}$ at 200 K are
shown as a function of the surface friction for a nonseparable
interaction potential \cite{ruth4} and two coverages (solid line
$\theta = 0.028$ and dashed line, $\theta = 0.18$). These jump rates
are obtained from Kramer's theory and squares and circles from
mean first passage time calculations. As can clearly seen, the left
part of the turnover region is slightly shifted depending on the
coverage used. Notice that without including finite barrier corrections,
the agreement is fairly good indicating that Kramers one dimensional
theory can be convenient for interpreting QHAS measurements even when
interacting adsorbates are present.

\begin{figure}
 \includegraphics[width=8cm]{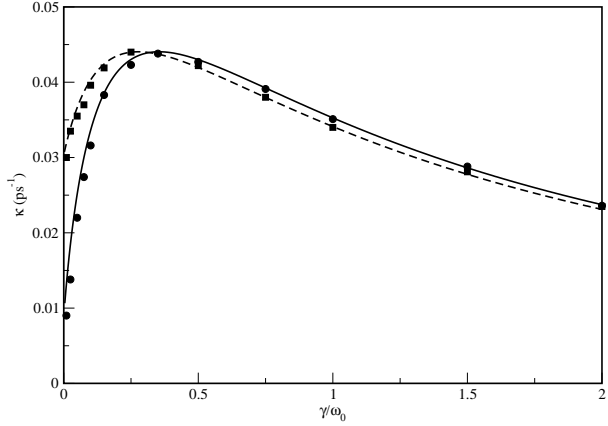}
 \caption{\label{fig1}
  Jump rates in ps$^{-1}$ as a function of the coverage (solid line
  $\theta = 0.028$; and dashed line, $\theta = 0.18$) for the
  diffusion of Na atoms on Cu(001) surface at 200 K, along the $[100]$.
  Kramers results are plotted in lines and mean first passage time
  calculations are plotted in circles and squares.}
\end{figure}

In Fig.~\ref{fig2} the tracer diffusion coefficient in a.u.
versus the total friction at 200~K is plotted. White circles are
issued from the ISA model and black circles from Kramer's results.
The agreement is again quite good and Einstein's law is clearly
fulfilled.

\begin{figure}
 \includegraphics[width=8cm]{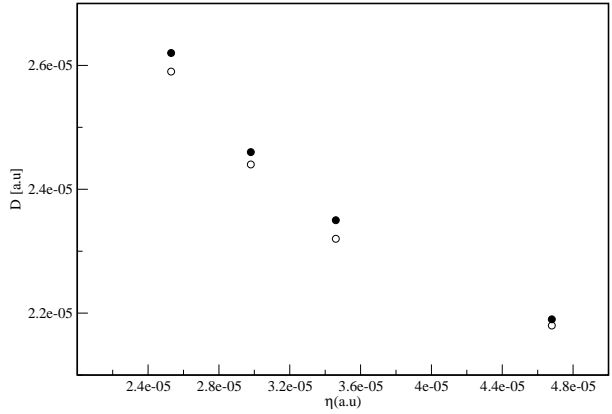}
 \caption{\label{fig2}
  Tracer diffusion coefficient in a. u. as a function of the
  coverage for the diffusion of Na atoms on Cu(001) surface
  at 200 K, along the $[100]$.
  Kramers results are given by black circles and Langevin
  numerical simulations by white circles.}
\end{figure}

Finally, in Fig.~\ref{fig3} the FWHM (in $\mu$eV) of the Q-peak is
shown for two different coverages (low, 0.028, and moderate, 0.18) at
200 K as a function of the parallel wave vector transfer covering the
first Brilloiun zone. Results from the ISA model are plotted in black
squares and circles. White squares and circles are experimental results
and the solid lines are coming from Kramer's theory which are also
obtained by assuming a Chudley-Elliott model. The agreement among
theoretical results is fairly good but with the experimental results are
poorer for the high coverage. Values of the coverage around 0.16 mark
the upper limit where the ISA model can work.

\begin{figure}
 \includegraphics[width=8cm]{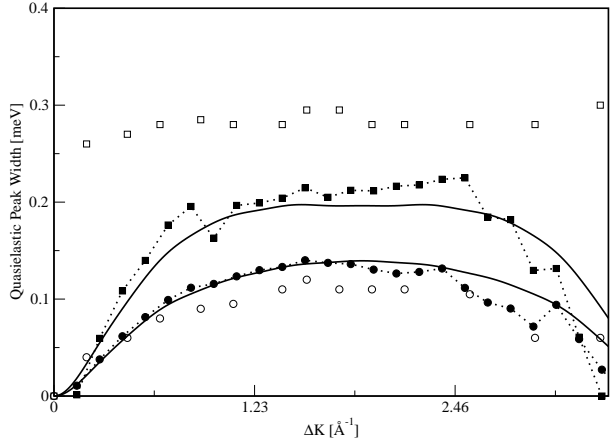}
 \caption{\label{fig3}
  Full width at half maximum (in $\mu$eV) of the quasielastic peak as a
  function of the parallel wave vector transfer for the diffusion of
  Na atoms on Cu(001) surface at 200 K and two coverages: 0.028 (low
  coverage) and 0.18 (intermediate coverage), along the $[100]$.}
\end{figure}


\subsubsection{Quantum dynamics}
\label{sec3.3.3}

Our starting point is again Eq.~(\ref{LME}) where now the adiabatic
force is derived by any general interaction potential. The same
discussion as in the harmonic case can be followed for the $I_1$
factor.

For Na atoms, the pairwise interaction potential is repulsive and
the mean interparticle distance should be most of the time greater than
$\lambda_B$. Thus, the $I_2$ factor could be replaced, in
a first approximation, by the classical counterpart, Eq.~(\ref{isfg}).
The error comes from small times but due to the fact
the diffusion process is a long time one, the influence on the
quasielastic peak (wave-vector dependence) and quantum diffusion
constant (Einstein's law) will be really small for massive particles
\cite{ruth5}.
In Fig.~\ref{fig4}, we show the effects of the quantum
correction in the diffusion process studied here at two different
surface temperatures for a coverage of 0.028. For comparison,
in this plot, the classical intermediate scattering function and the
real part of its quantum analog are displayed.
As seen, although the Na atom is a relatively massive particle, at low
temperatures the {\it plateau} is lower for the quantum case.
This implies an initially faster decay arising from the
strong influence of the quantum behavior at short time scales.
It is therefore the real part of the intermediate scattering function
what one should compare to the experiment rather than $I_2$, as
is usually done. Obviously, this effect will be less pronounced at
high coverages.

\begin{figure}
 \includegraphics[width=8cm]{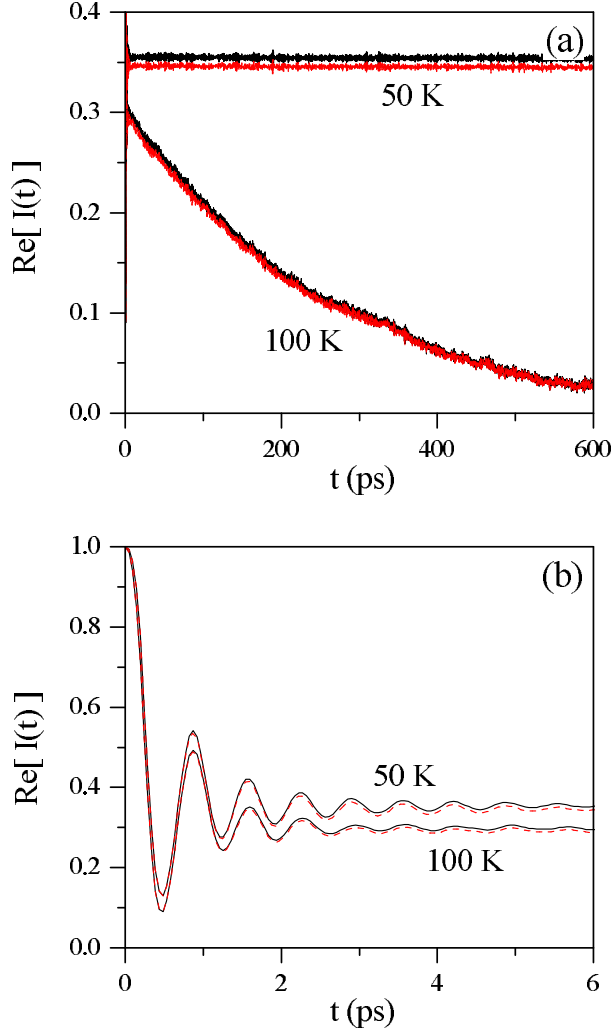}
 \caption{\label{fig4}
  (Color online.)
  Classical intermediate scattering function (\ref{eq8}) for Na
  diffusion on a Cu(001) surface at 50~K and 100~K (black solid lines)
  and the real part of its quantum-mechanical analog (\ref{eq6})
  (red/dark grey dashed lines).
  The surface coverage considered here is 0.028 and the diffusion
  along the azimut $[100]$.}
\end{figure}

At very low temperatures, the Matsubara series should play a
similar role like the extra term observed in the $f$ and $g$
functions like in the harmonic case. Thus, a new quantum correction
should be added for very low temperatures.


\section{Conclusions}
\label{sec4}

It is remarkable that, within the Markovian formalism presented here,
the quantum intermediate scattering function, $I_1$, is independent
of the relative corrugation of the surface and, at short times, also
independent of the friction.
At low surface temperatures, the $I_1$ factor will be responsible
for a higher contribution of the imaginary part of $I$, given by
Eq.~(\ref{eq6}), modifying substantially the response in the diffusion
process. For relatively heavy particles and at very long times
(diffusion time scales), operators in the $I_2$ factor can be
replaced by variables, since $\lambda_B$ is very small.
As far as we know, an exact quantum calculation for a corrugated
surface is not possible and some approximations have to be invoked,
e.g., the damped harmonic oscillator  has been applied here
in order to obtain close formulas.
Of course, other different, alternative theoretical approaches
can also be found in literature (see, for example,
Refs.~\onlinecite{peter,others}) but within the single adsorbate
approximation.
The theoretical formalism that we propose here should also be very
useful to avoid extrapolations at zero surface temperature when trying
to extract information about the frustrated translational mode.
Diffusion experiments at low temperatures are very difficult to
perform (or even unaffordable); for example, for the HeSE technique
surface temperatures around 100 K can be attainable.
However, the type of theoretical calculations needed in this formalism
is easy to carry out and they would provide a simple manner to go to
lower temperatures with quite reliable results, thus allowing to
extract confident values of magnitudes such as friction coefficients
and oscillation frequencies.
By decreasing the surface temperature, quantum effects are extended
at higher values of time. Going from 100 K to 50 K, the time where
the quantum dynamics is important increases from 0.07 ps to 0.15 ps.
The standard propagation time for diffusion is greater than 400 ps.
In our opinion, the limits of applicability of this quantum theory
should be  at coverages up to 12-16 $\%$ and around 50 K where the
Matsubara series is still playing no role on the diffusion. Obviously,
at lower surface temperatures, the quantum character of the noise
becomes more and more important. This type of conditions as well as
the diffusion mediated by tunneling needs to be more investigated
since new experimental results are being analyzed.


\section*{Acknowledgements}

We would like to take advantage of this opportunity to express our
deep gratitude to Prof.\ Eli Pollak, a teacher and a reference at
scientific level and a close friend at  personal level. Thanks Eli.

This work has been supported by the Ministerio de Ciencia e
Innovaci\'on (Spain) under Projects FIS2007-02461 and SB2006-0011
(G.R.L.).
R.M.-C.\ thanks the Royal Society for a Newton Fellowship; A.S.\ Sanz
thanks the Consejo Superior de Investigaciones Cient\'{\i}ficas for a
JAE-Doc contract.



\begin{thebibliography}{99}
\eprint{}

\bibitem{gordon}
 R. G. Gordon,  Adv. Mag. Resonance 3 (1968) 1.

\bibitem{mcquarrie}
 D. A. McQuarrie, Statistical Mechanics, Harper and Row, New York, 1976.

\bibitem{kubo}
 R. Kubo, Rep. Prog. Phys. 29 (1966) 255.

\bibitem{vanhove}
 L. Van Hove,  Phys. Rev. 95 (1954) 249.

\bibitem{vineyard}
 G. H. Vineyard, Phys. Rev. 110 (1958) 999.

\bibitem{lovesey}
 S. W. Lovesey, Theory of Neutron Scattering from Condensed
 Matter, Clarendon, Oxford, 1984.

\bibitem{hansen}
 J. P. Hansen, I. R. McDonald, Theory of Simple Liquids,
 Academic Press, London, 1986.

\bibitem{bee}
 M. B\'ee, Quasielastic Neutron Scattering,
 Adam Hilger, Bristol, 1988.

\bibitem{lahee}
 A. M. Lahee, J. R. Manson, J. P. Toennies, Ch. W\"oll,
 Phys. Rev. Lett. 57 (1986) 471; {\it ibid}
 J. Chem. Phys. 86 (1987) 7194.

\bibitem{manson}
 J. R. Manson, V. Celli, Phys. Rev. B 39 (1989) 3605.

\bibitem{hofmann}
 F. Hofmann, J. P. Toennies, Chem. Rev. 78 (1996) 3900.

\bibitem{eli1}
 S. Miret-Art\'es, E. Pollak, J. Phys.: Condens. Matter 17 (2005) S4133.

\bibitem{graham}
 A. P. Graham, F. Hofmann, J. P. Toennies, L. Y. Chen, S. C. Ying,
 Phys. Rev. B 56 (1997) 10567.

\bibitem{elis}
 J. Ellis, A. P. Graham, F. Hofmann, J. P. Toennies,
 Phys. Rev. B 63 (2001) 195408.

\bibitem{vega1}
 J. L. Vega, R. Guantes, S. Miret-Art\'es,
 J. Phys.: Condens. Matter 14 (2002) 6193.

\bibitem{vega2}
 J. L. Vega, R. Guantes, S. Miret-Art\'es,
 J. Phys.: Condens. Matter  16 (2004) S2879.


\bibitem{vega3}
 J. L. Vega, R. Guantes, S. Miret-Art\'es, Phys. Chem.
 Chem. Phys. 4 (2002) 4985.

\bibitem{guantes1}
 R. Guantes, J. L. Vega, S. Miret-Art\'es, E. Pollak,
 J. Chem. Phys. 119 (2003) 2780.

\bibitem{guantes2}
 R. Guantes, J. L. Vega, S. Miret-Art\'es, E. Pollak,
 J. Chem. Phys. 120 (2004) 10768.

\bibitem{ruth1}
 R. Mart\'{\i}nez-Casado, J. L. Vega, A. S. Sanz, S. Miret-Art\'es,
 Phys. Rev. Lett. 98 (2007) 216102.

\bibitem{ruth2}
 R. Mart\'{\i}nez-Casado, J. L. Vega, A. S. Sanz, S. Miret-Art\'es,
 Phys. Rev. E 75 (2007) 051128.

\bibitem{ruth3}
 R. Mart\'{\i}nez-Casado, J. L. Vega, A. S. Sanz, S. Miret-Art\'es,
 J. Phys.: Condens. Matter 19 (2007) 305002.

\bibitem{ruth4}
 R. Mart\'{\i}nez-Casado, J. L. Vega, A. S. Sanz, S. Miret-Art\'es,
 Phys. Rev. B 77 (2008) 115414.

\bibitem{vvleck-weiss}
 J. H. van Vleck, V. F. Weisskopf, Rev. Mod. Phys. 17 (1945) 227.

\bibitem{gardiner}
 C. W. Gardiner, Handbook of Stochastic Methods,
 Springer-Verlag, Berlin, 1983.

\bibitem{ellis}
 J. Ellis and J. P. Toennies, Phys. Rev. B 56 (1997) 15378.


\bibitem{allison0}
 H. Hedgeland, P. Fouquet, A. P. Jardine, G. Alexandrowicz, W. Allison,
 J. Ellis, Nature Phys. 5 (2009) 561.

\bibitem{german}
 R. Mart\'{\i}nez-Casado, A. S. Sanz, G. Rojas-Lorenzo,
 S. Miret-Art\'es, J. Chem. Phys. 132 (2010) 054704.


\bibitem{allison1}
 G. Alexandrowicz, A. P. Jardine, H. Hedgeland, W. Allison,
 J. Ellis, Phys. Rev. Lett. 97 (2006) 156103.

\bibitem{allison2}
 A. P. Jardine, G. Alexandrowicz, H. Hedgeland, R.D. Diehl, W. Allison,
 J. Ellis, J. Phys.: Condens. Matter 19 (2007) 305010.

\bibitem{allison3}
 A. P. Jardine, H. Hedgeland, G. Alexandrowicz, W. Allison,
 J. Ellis, Prog. Surf. Sci. 84 (2009) 323.


\bibitem{farago}
 B. Farago, Physica B 267-268 (1999) 270.

\bibitem{fouquet}
 P. Fouquet, H. Hedgeland, A. Jardine, G. Alexandrowicz, W. Allison,
 J. Ellis, Physica B 385-386 (2006) 269.

\bibitem{ruth5}
 R. Mart\'{\i}nez-Casado, A. S. Sanz, S. Miret-Art\'es,
 J. Chem. Phys. 129 (2008) 184704.

\bibitem{yip91} J. P. Boon, S. Yip, Molecular Hydrodynamics,
 Dover Publications, New York, 1991.

\bibitem{schofield}
 P. Schofield, Phys. Rev. Lett. 4 (1960) 239.

\bibitem{maga}
 V. B. Magalinskii, Sov. Phys. JETP 9 (1959) 1381.

\bibitem{caldeira}
 A. O. Caldeira, A. J. Leggett, Ann. Phys. (N. Y.) 149 (1983) 374;
 ibid. 153 (1984) 445.

\bibitem{weiss}
 U. Weiss, Quantum Dissipative Systems,
 World Scientific, Singapore, 1993.


\bibitem{ingold}
 G. Ingold, Lec. Notes Phys. 611 (2002) 1.

\bibitem{david}
 R. Guantes, J. L. Vega, S. Miret-Art\'es, D. A. Micha,
 J. Chem. Phys. 120 (2004) 10768.

\bibitem{jcp-elliot}
 R. Mart\'{\i}nez-Casado, J. L. Vega, A. S. Sanz, S. Miret-Art\'es,
 J. Chem. Phys. 126 (2007) 194711.

\bibitem{toeprl}
 J. Ellis, A. P. Graham, J. P. Toennies,
 Phys. Rev. Lett. 82 (1999) 5072.

\bibitem{ruth6}
 R. Mart\'{\i}nez-Casado, J. L. Vega, A. S. Sanz, S. Miret-Art\'es,
 J. Phys.: Condens. Matter 19 (2007) 176006.

\bibitem{risken}
 H. Risken, The Fokker-Planck Equation, Springer-Verlag, Berlin, 1984.

\bibitem{elliott}
 C. T. Chudley, R. J. Elliott, Proc. Phys. Soc. 57 (1961) 353.

\bibitem{melnikov}
 V. I. Mel'nikov, S. V. Meshkov, J. Chem. Phys. 85 (1986) 1018;
 V. I. Mel'nikov, Phys. Rep. 209 (1991) 1.

\bibitem{eli3}
 E. Pollak, J. Chem. Phys. 85 (1986) 865.
 Y. Georgievskii, E. Pollak, Phys. Rev. E 49 (1994) 5098:
 Y. Georgievskii, M. A. Kozhushner, E. Pollak,
 J. Chem. Phys. 102 (1995) 6908.

\bibitem{kramers}
 H. A. Kramers, Physica (Utrecht) 7 (1940) 284.

\bibitem{peter}
 E. Pollak, H. Grabert, P. H\"anggi, J. Chem. Phys. 91 (1989) 4037;
 P. H\"anggi, P. Talkner, M. Borkovec, Rev. Mod. Phys. 62 (1990) 251.

\bibitem{grote}
 R. F. Grote, J. T. Hynes, J. Chem. Phys. 73 (1980) 2715.

\bibitem{allen}
 M.P. Allen, D.J. Tildesley, Computer Simulation of Liquids,
 Clarendon, Oxford, 1990.

\bibitem{others}
 L.Y. Chen, S.C. Ying, Phys. Rev. Lett. 73 (1994) 700;
 Y. Georgievskii, E. Pollak, Phys. Rev. E 49 (1994) 5098.

\end{thebibliography}
\end{document}